\documentclass[journal]{IEEEtran}
\usepackage{graphicx,amsmath,amssymb,color,bm,subfigure}
\usepackage{algorithm2e}
\DeclareMathOperator{\rank}{rank}

\DeclareMathOperator{\tmid}{mid}
\DeclareMathOperator{\TB}{TB}
\DeclareMathOperator{\TG}{TG}
\DeclareMathOperator{\GTG}{GTG}
\DeclareMathOperator{\GM}{GM}

\begin{document}
\newtheorem{numThm}{Theorem}
\newtheorem{numCor}{Corollary}
\newtheorem{numLem}{Lemma}
\newtheorem{numProp}{Proposition}
\newtheorem{numDef}{Definition}
\newtheorem{numClaim}{Claim}
\newtheorem{numConj}{Conjecture}
\newtheorem{numRem}{Remark}
\newtheorem{numProb}{Problem}
\renewcommand{\QED}{\QEDopen}
\title{The Extraction and Complexity Limits of\\Graphical Models for Linear Codes}
\author{\authorblockN{Thomas R. Halford and Keith M. Chugg}\\
\authorblockA{Communication Sciences Institute\\
Department of Electrical Engineering - Systems\\
University of Southern California\\
Los Angeles, CA 90089-2565\\
Email: \{halford,chugg\}@usc.edu}}
\maketitle
\begin{abstract}
Two broad classes of graphical modeling problems for codes can be identified in the literature: constructive and extractive problems.  The former class of problems concern the \textit{construction} of a graphical model in order to define a new code.  The latter class of problems concern the \textit{extraction} of a graphical model for a (fixed) given code.  The design of a new low-density parity-check code for some given criteria (e.g. target block length and code rate) is an example of a constructive problem.  The determination of a graphical model  for a classical linear block code which implies a decoding algorithm with desired performance and complexity characteristics is an example of an extractive problem.  This work focuses on extractive graphical model problems and aims to lay out some of the foundations of the theory of such problems for linear codes.

The primary focus of this work is a study of the space of all graphical models for a (fixed) given code.  The tradeoff between cyclic topology and complexity in this space is characterized via the introduction of a new bound: the tree-inducing cut-set bound.  The proposed bound provides a more precise characterization of this tradeoff than that which can be obtained using existing tools (e.g. the Cut-Set Bound) and can be viewed as a generalization of the square-root bound for tail-biting trellises to graphical models with arbitrary cyclic topologies.  Searching the space of graphical models for a given code is then enabled by introducing a set of basic graphical model transformation operations  which are shown to span this space.  Finally, heuristics for extracting novel graphical models for linear block codes using these transformations are investigated.
\end{abstract}
\section{Introduction}\label{intro_sec}
Graphical models of codes have been studied since the 1960s and this study has intensified in recent years due to the discovery of turbo codes by Berrou \textit{et al.} \cite{BeGlTh93}, the rediscovery of Gallager's low-density parity-check (LDPC) codes \cite{Ga62} by Spielman \textit{et al.} \cite{SiSp96} and MacKay \textit{et al.} \cite{Ma97}, and the pioneering work of Wiberg, Loeliger and Koetter \cite{WiLoKo95b,Wi96}.  It is now well-known that together with a suitable message passing schedule, a graphical model implies a soft-in soft-out (SISO) decoding algorithm which is optimal for cycle-free models and suboptimal, yet often substantially less complex, for cyclic models (cf. \cite{Wi96,AjMc00,ChAnCh01,KsFrLo01,Fo01}).  It has been observed empirically in the literature that there exists a correlation between the cyclic topology of a graphical model and the performance of the decoding algorithms implied by that graphical model (cf. \cite{WiLoKo95b,Fo01,Ta81,MaBa01,ArElHu01,GeEpSm01,TiJoViWe04,HaCh06}).  To summarize this empirical ``folk-knowledge'', those graphical models which imply near-optimal decoding algorithms tend to have large girth, a small number of short cycles and a cycle structure that is not overly regular.

Two broad classes of graphical modeling problems can be identified in the literature:
\begin{itemize}
\item \textit{Constructive} problems: Given a set of design requirements, design a suitable code by constructing a good graphical model (i.e. a model which implies a low-complexity, near-optimal decoding algorithm).
\item \textit{Extractive} problems: Given a specific (fixed) code, extract a graphical model for that code which implies a decoding algorithm with desired complexity and performance characteristics.
\end{itemize}
Constructive graphical modeling problems have been widely addressed by the coding theory community.  Capacity approaching LDPC codes have been designed for both the additive white Gaussian noise (AWGN) channel (cf. \cite{ChFoRiUr01,RiShUr01}) and the binary erasure channel (cf. \cite{OsSh02,PfSaUr05}).  Other classes of modern codes have been successfully designed for a wide range of practically motivated block lengths and rates (cf. \cite{BeMo96,DoDiPo98,Be03,Br01,ChThDiGrMe05}).

Less is understood about extractive graphical modeling problems, however.  The extractive problems that have received the most attention are those concerning Tanner graph \cite{Ta81} and trellis representations of block codes.  Tanner graphs imply low-complexity decoding algorithms; however, the Tanner graphs corresponding to many block codes of practical interest, e.g. high-rate Reed-Muller (RM), Reed-Solomon (RS), and Bose-Chaudhuri-Hocquenghem (BCH) codes, necessarily contain many short cycles \cite{HaGrCh06} and thus imply poorly performing decoding algorithms.  There is a well-developed theory of conventional trellises \cite{Va99} and tail-biting trellises \cite{CaFoVa99,KoVa03c} for linear block codes.  Conventional and tail-biting trellises imply optimal and, respectively, near-optimal decoding algorithms; however, for many block codes of practical interest these decoding algorithms are prohibitively complex thus motivating the study of more general graphical models (i.e. models with a richer cyclic topology than a single cycle).  

The goal of this work is to lay out some of the foundations of the theory of extractive graphical modeling problems.  Following a review of graphical models for codes in Section \ref{background_sec}, a complexity measure for graphical models is introduced in Section \ref{complexity_sec}.  The proposed measure captures a cyclic graphical model analog of the familiar notions of state and branch complexity for trellises \cite{Va99}.  The \textit{minimal tree complexity} of a code, which is a natural generalization of the well-understood minimal trellis complexity of a code to arbitrary cycle-free models, is then defined using this measure.

The tradeoff between cyclic topology and complexity in graphical models is studied in Section \ref{ticsb_sec}.   Wiberg's Cut-Set Bound (CSB) is the existing tool that best characterizes this fundamental tradeoff \cite{Wi96}.  While the CSB can be used to establish the square-root bound for tail-biting trellises \cite{CaFoVa99} and thus provides a precise characterization of the potential tradeoff between cyclic topology and complexity for single-cycle models, as was first noted by Wiberg \textit{et al.} \cite{WiLoKo95b}, it is very challenging to use the CSB to characterize this tradeoff for graphical models with cyclic topologies richer than a single cycle.  In order to provide a more precise characterization of this tradeoff than that offered by the CSB alone, this work introduces a new bound in Section \ref{ticsb_sec} - the \textit{tree-inducing cut-set bound} - which may be viewed as a generalization of the square-root bound to graphical models with arbitrary cyclic topologies.  Specifically, it is shown that an $r^{th}$-root complexity reduction (with respect to the minimal tree complexity as defined in Section \ref{complexity_sec}) requires the introduction of \textit{at least} $r(r-1)/2$ cycles.  The proposed bound can thus be viewed as an extension of the square-root bound to graphical models with arbitrary cyclic topologies.

The transformation of graphical models is studied in Section \ref{model_tx_sec} and \ref{extraction_sec}.  Whereas minimal conventional and tail-biting trellis models can be characterized algebraically via trellis-oriented generator matrices \cite{Va99}, there is in general no known analog of such algebraic characterizations for arbitrary cycle-free graphical models \cite{Fo03}, let alone cyclic models.  In the absence of such an algebraic characterization, it is initially unclear as to how cyclic graphical models can be extracted.  In Section \ref{model_tx_sec}, a set of basic transformation operations on graphical models for codes is introduced and it is shown that any graphical model for a given code can be transformed into any other graphical model for that same code via the application of a finite number of these basic transformations.  The transformations studied in Section \ref{model_tx_sec} thus provide a mechanism for searching the space of all \textit{all} graphical models for a given code.  In Section \ref{extraction_sec}, the basic transformations introduced in Section \ref{model_tx_sec} are used to extract novel graphical models for linear block codes.  Starting with an initial Tanner graph for a given code, heuristics for extracting other Tanner graphs, generalized Tanner graphs, and more complex cyclic graphical models are investigated.  Concluding remarks and directions for future work are given in Section \ref{conc_sec}.
\section{Background}\label{background_sec}
\subsection{Notation}\label{notation_subsec}
The binomial coefficient is denoted $\binom{a}{b}$ where $a,b\in\mathbb{Z}$ are integers.  The finite field with $q$ elements is denoted $\mathbb{F}_q$.  Given a finite index set $I$, the vector space over $\mathbb{F}_q$ defined on $I$ is the set of vectors
\begin{equation}
\mathbb{F}_q^I=\left\{\pmb{f}=\left(f_i\in\mathbb{F}_q,i\in I\right)\right\}.
\end{equation}
Suppose that $J\subseteq I$ is some subset of the index set $I$.  The \textit{projection} of a vector $\pmb{f}\in\mathbb{F}_q^I$ onto $J$ is denoted
\begin{equation}
\pmb{f}_{|J}=\left(f_i,i\in J\right).
\end{equation}
\subsection{Codes, Projections, and Subcodes}\label{codes_subsec}
Given a finite index set $I$, a \textit{linear code} over $\mathbb{F}_q$ defined on $I$ is some vector subspace $\mathcal{C}\subseteq\mathbb{F}_q^I$.  The \textit{block length} and \textit{dimension} of $\mathcal{C}$ are denoted $n(\mathcal{C})=\left|I\right|$ and $k(\mathcal{C})=\dim\mathcal{C}$, respectively.  If known, the minimum Hamming distance of $\mathcal{C}$ is denoted $d(\mathcal{C})$ and $\mathcal{C}$ may be described by the triplet $[n(\mathcal{C}),k(\mathcal{C}),d(\mathcal{C})]$.  This work considers only linear codes and  the terms code and linear code are used interchangeably. 

A code $\mathcal{C}$ can be described by an $r_G\times n(\mathcal{C})$, $r_G\geq k(\mathcal{C})$, \textit{generator matrix} $G_\mathcal{C}$ over $\mathbb{F}_q$, the rows of which span $\mathcal{C}$.  An $r_G\times n(\mathcal{C})$ generator matrix is \textit{redundant} if $r_G$ is strictly greater than $k(\mathcal{C})$.  A code $\mathcal{C}$ can also be described by an $r_H\times n(\mathcal{C})$, $r_H\geq n(\mathcal{C})-k(\mathcal{C})$, \textit{parity-check matrix} $H_\mathcal{C}$ over $\mathbb{F}_q$, the rows of which span the null space of $\mathcal{C}$ (i.e. the dual code $\mathcal{C}^\perp$).  Each row of $H_\mathcal{C}$ defines a $q$\textit{-ary single parity-check equation} which every codeword in $\mathcal{C}$ must satisfy.  An $r_H\times n(\mathcal{C})$ parity-check matrix is \textit{redundant} if $r_H$ is strictly greater than
\begin{equation}
k(\mathcal{C}^\perp)=n(\mathcal{C})-k(\mathcal{C}).
\end{equation}  

Given a subset $J\subseteq I$ of the index set $I$, the \textit{projection} of $\mathcal{C}$ onto $J$ is the set of all codeword projections: 
\begin{equation}
\mathcal{C}_{|J}=\left\{\pmb{c}_{|J},\pmb{c}\in\mathcal{C}\right\}.
\end{equation}
Closely related to $\mathcal{C}_{|J}$ is the \textit{subcode} $\mathcal{C}_J$: the projection onto $J$ of the subset of codewords satisfying $c_i=0$ for $i\in I\setminus J$.  Both $\mathcal{C}_{|J}$ and $\mathcal{C}_{J}$ are linear codes.  

Suppose that $\mathcal{C}_1$ and $\mathcal{C}_2$ are two codes over $\mathbb{F}_q$ defined on the same index set $I$.  The intersection $\mathcal{C}_1\cap\mathcal{C}_2$ of $\mathcal{C}_1$ and $\mathcal{C}_2$ is a linear code defined on $I$ comprising the vectors in $\mathbb{F}_q^I$ that are contained in both $\mathcal{C}_1$ and $\mathcal{C}_2$.

Finally, suppose that $\mathcal{C}_a$ and $\mathcal{C}_b$ are two codes defined on the disjoint index sets $J_a$ and $J_b$, respectively.  The Cartesian product $\mathcal{C}=\mathcal{C}_a\times\mathcal{C}_b$ is the code defined on the index set $J=\left\{J_a,J_b\right\}$ such that $\mathcal{C}_{|J_a}=\mathcal{C}_{J_a}=\mathcal{C}_a$ and $\mathcal{C}_{|J_b}=\mathcal{C}_{J_b}=\mathcal{C}_b$.
\subsection{Generalized Extension Codes}
Let $\mathcal{C}$ be a linear code over $\mathbb{F}_q$ defined on the index set $I$.  Let $J\subseteq I$ be some subset of $I$ and let
\begin{equation}
\pmb\beta=(\beta_j\neq 0\in\mathbb{F}_q,j\in J)
\end{equation}
be a vector of non-zero elements of $\mathbb{F}_q$.  A \textit{generalized extension} of $\mathcal{C}$ is formed by adding a $q$-ary parity-check on the codeword coordinates indexed by $J$ to $\mathcal{C}$ (i.e. a $q$-ary partial parity symbol).  The generalized extension code $\widetilde{\mathcal{C}}$ is defined on the index set $\widetilde{I}=I\cup\{p\}$ such that if $\pmb{c}=(c_i,i\in I)\in\mathcal{C}$ then $\widetilde{\pmb{c}}=(\widetilde{c}_i,i\in \widetilde{I})\in\widetilde{\mathcal{C}}$ where $\widetilde{c}_i=c_i$ if $i\in I$ and
\begin{equation}
\widetilde{c}_p=\sum_{j\in J}\beta_jc_j.
\end{equation}
The length and dimension of $\widetilde{\mathcal{C}}$ are $n(\widetilde{\mathcal{C}})=n(\mathcal{C})+1$ and $k(\widetilde{\mathcal{C}})=k(\mathcal{C})$, respectively, and the minimum distance of $\widetilde{\mathcal{C}}$ satisfies $d(\widetilde{\mathcal{C}})\in\left\{d(\mathcal{C}),d(\mathcal{C})+1\right\}$.  Note that if $J=I$ and $\beta_j=1$ for all $j\in J$, then $\widetilde{\mathcal{C}}$ is simply a classically defined extended code \cite{MaSl78}.  More generally, a \textit{degree-}$g$ \textit{generalized extension} of $\mathcal{C}$ is formed by adding $g$ $q$-ary partial parity symbols to $\mathcal{C}$ and is defined on the index set $I\cup \{p_1,p_2,\ldots,p_g\}$.  The $j^{th}$ partial parity symbol $c_{p_j}$ in such an extension is defined as a partial parity on some subset of $I\cup \{p_1,\ldots,p_{j-1}\}$. 
\subsection{Graph Theory}\label{graph_theory_subsec}
A \textit{graph} $\mathcal{G}=\left(\mathcal{V},\mathcal{E},\mathcal{H}\right)$ consists of:
\begin{itemize} 
\item A finite non-empty set of vertices $\mathcal{V}$.
\item A set of edges $\mathcal{E}$, which is some subset of the pairs $\{\{u,v\}:
u,v\in\mathcal{V},u\neq v\}$.
\item A set of \textit{half-edges} $\mathcal{H}$, which is any subset of $\mathcal{V}$.
\end{itemize}
It is non-standard to define graphs with half-edges; however, as will be demonstrated in Section \ref{graphical_models_subsec}, half-edges are useful in the context of graphical models for codes.  A walk of length $n$ in $\mathcal{G}$ is a sequence of vertices $v_1,v_2,\cdots,v_n,v_{n+1}$ in $\mathcal{V}$ such that $\{v_i,v_{i+1}\}\in\mathcal{E}$ for all $i\in\left\{1,\ldots,n\right\}$.  A path is a walk on distinct vertices while a \textit{cycle} of length $n$ is a walk such that $v_1$ through $v_n$ are distinct and $v_1=v_{n+1}$.  Cycles of length $n$ are often denoted $n$-cycles.  A \textit{tree} is a graph containing no cycles (i.e. a \textit{cycle-free} graph).  Two vertices $u,v\in\mathcal{V}$ are \textit{adjacent} if a single edge $\{u,v\}\in\mathcal{E}$ connects $u$ to $v$.  A graph is \textit{connected} if any two of its vertices are linked by a walk.   A \textit{cut} in a connected graph $\mathcal{G}$ is some subset of edges $\mathcal{X}\subseteq\mathcal{E}$ the removal of which yields a disconnected graph.  Cuts thus partition the vertex set $\mathcal{V}$.  Finally, a graph is bipartite if its vertex set can be partitioned $\mathcal{V}=\mathcal{U}\cup\mathcal{W}$, $\mathcal{U}\cap\mathcal{W}=\emptyset$, such that any edge in $\mathcal{E}$ joins a vertex in $\mathcal{U}$ to one in $\mathcal{W}$.
\subsection{Graphical Models of Codes}\label{graphical_models_subsec}
Graphical models for codes have been described by a number of different authors using a wide variety of notation (e.g. \cite{Wi96,AjMc00,ChAnCh01,KsFrLo01,Fo01,Ta81}).  The present work uses the notation described below which was established by Forney in his \textit{Codes on Graphs} papers \cite{Fo01,Fo03}.

A \textit{linear behavioral realization} of a linear code $\mathcal{C}\subseteq\mathbb{F}_q^I$ comprises three sets:
\begin{itemize}
\item A set of \textit{visible} (or symbol) variables $\big\{V_i,i\in I\big\}$  corresponding to the codeword coordinates\footnote{Observe that this definition is slightly different than that proposed in \cite{Fo03} which permitted the use of $q^r$-ary visible variables corresponding to $r$ codeword coordinates.  By appropriately introducing equality constraints and $q$-ary hidden variables, it can be seen that these two definitions are essentially equivalent.} with alphabets $\big\{\mathbb{F}_q^i,i\in I\big\}$.
\item A set of \textit{hidden} (or state) variables $\big\{S_i,i\in I_S\big\}$ with alphabets $\big\{\mathbb{F}_q^{T_i},i\in I_S\big\}$.
\item A set of linear local constraint codes $\big\{\mathcal{C}_i,i\in I_C\big\}$.
\end{itemize}
Each visible variable is $q$-ary while the hidden variable $S_i$ with alphabet $\mathbb{F}_q^{T_i}$ is $q^{|T_i|}$-ary.  The hidden variable alphabet index sets $\big\{T_i,i\in I_S\big\}$ are disjoint and unrelated to $I$.  Each local constraint code $\mathcal{C}_i$ involves a certain subset of the visible, $I_V(i)\subseteq I$, and hidden, $I_S(i)\subseteq I_S$, variables and defines a subspace of the local configuration space:
\begin{equation}
\mathcal{C}_i\subseteq\Bigg(\prod_{j\in I_V(i)}\mathbb{F}_q^{j}\Bigg)\times\Bigg(\prod_{j\in I_S(i)}\mathbb{F}_q^{T_j}\Bigg).
\end{equation}
Each local constraint code $\mathcal{C}_i$ thus has a well-defined block length 
\begin{equation}
n(\mathcal{C}_i)=\left|I_V(i)\right|+\sum_{j\in I_S(i)}\left|T_j\right|
\end{equation}
and dimension $k(\mathcal{C}_i)=\dim\mathcal{C}_i$ over $\mathbb{F}_q$.  Local constraints that involve only hidden variables are \textit{internal} constraints while those involving visible variables are \textit{interface} constraints.  The full behavior of the realization is the set $\mathfrak{B}$ of all visible and hidden variable configurations which simultaneously satisfy all local constraint codes:
\begin{equation}
\mathfrak{B}\subseteq \Bigg(\prod_{i\in I}\mathbb{F}_q^{i}\Bigg)\times\Bigg(\prod_{j\in I_S}\mathbb{F}_q^{T_j}\Bigg)=\mathbb{F}_q^I\times\Bigg(\prod_{j\in I_S}\mathbb{F}_q^{T_j}\Bigg).
\end{equation}
The projection of the linear code $\mathfrak{B}$ onto $I$ is precisely $\mathcal{C}$. 

Forney demonstrated in \cite{Fo01} that it is sufficient to consider only those realizations in which all visible variables are involved in a single local constraint and all hidden variables are involved in two local constraints.  Such \textit{normal} realizations have a natural graphical representation in which local constraints are represented by vertices, visible variables by half-edges and hidden variables by edges.  The half-edge corresponding to the visible variable $V_i$ is incident on the vertex corresponding to the single local constraint which involves $V_i$.  The edge corresponding to the hidden variable $S_j$ is incident on the vertices corresponding to the two local constraints which involve $S_j$.  The notation $\mathcal{G}_\mathcal{C}$ and term \textit{graphical model} is used throughout  this work to denote both a normal realization of a code $\mathcal{C}$ and its associated graphical representation.

It is assumed throughout that the graphical models considered are connected.  Equivalently, it is assumed throughout that the codes studied cannot be decomposed into Cartesian products of shorter codes \cite{Fo01}.  Note that this restriction will apply only to the global code considered and not to the local constraints in a given graphical model.
\subsection{Tanner Graphs and Generalized Tanner Graphs}\label{tg_gtg_subsec}
The term \textit{Tanner graph} has been used to describe different classes of graphical models by different authors.  Tanner graphs denote those graphical models corresponding to parity-check matrices in this work.  Specifically, let $H_\mathcal{C}$ be an $r_H\times n(\mathcal{C})$ parity-check matrix for the code $\mathcal{C}$ over $\mathbb{F}_q$ defined on the index set $I$.  The Tanner graph corresponding to $H_\mathcal{C}$ contains $r_H+n(\mathcal{C})$ local constraints of which $n(\mathcal{C})$ are interface repetition constraints, one corresponding to each codeword coordinate, and $r_H$ are internal $q$-ary single parity-check constraints, one corresponding to each row of $H_\mathcal{C}$.  An edge (hidden variable) connects a repetition constraint $\mathcal{C}_i$ to a single parity-check constraint $\mathcal{C}_j$ if and only if the codeword coordinate corresponding to $\mathcal{C}_i$ is involved in the single parity-check equation defined by the row corresponding to $\mathcal{C}_j$.  A Tanner graph for $\mathcal{C}$ is \textit{redundant} if it corresponds to a redundant parity-check matrix.  A \textit{degree}-$g$ \textit{generalized Tanner graph} for $\mathcal{C}$ is simply a Tanner graph corresponding to some degree-$g$ generalized extension of $\mathcal{C}$ in which the visible variables corresponding to the partial parity symbols have been removed.  Generalized Tanner graphs have been studied previously in the literature under the rubric of generalized parity-check matrices \cite{Ma00,YeChFo02}.
\section{A Complexity Measure for Graphical Models}\label{complexity_sec}
\subsection{$q^m$-ary Graphical Models}\label{qmary_subsec}
This work introduces the term $q^m$\textit{-ary graphical model} to denote a normal realization of a linear code $\mathcal{C}$ over $\mathbb{F}_q$ that satisfies the following constraints:
\begin{itemize}
\item The alphabet index size of every hidden variable\\
$S_i$, $i\in I_S$, satisfies $|T_i|\leq m$.
\item Every local constraint $\mathcal{C}_i$, $i\in I_C$, either satifies
\begin{equation}
\min\left(k(\mathcal{C}_i),n(\mathcal{C}_i)-k(\mathcal{C}_i)\right)\leq m
\end{equation}
or can be decomposed as a Cartesian product of\\ 
codes, each of which satisfies this condition.
\end{itemize}
The complexity measure $m$ simultaneously captures a cyclic graphical model analog of the familiar notions of state and branch complexity for trellises \cite{Va99}.  From the above definition, it is clear that Tanner graphs and generalized Tanner graphs for codes over $\mathbb{F}_q$ are $q$-ary graphical models.  The efficacy of this complexity measure is discussed further in Section \ref{ti_csb_discussion_subsec}.
\subsection{Properties of $q^m$-ary Graphical Models}\label{model_prop_subsec}
The following three properties of $q^m$-ary graphical models will be used in the proof of Theorem \ref{ti_csb_theorem} in Section \ref{ticsb_sec}:
\begin{enumerate}
\item \textit{Internal Local Constraint Involvement Property:} Any hidden variable in a $q^m$-ary graphical model can be made to be incident on an \textit{internal} local constraint $\mathcal{C}_i$ which satisfies $n(\mathcal{C}_i)-k(\mathcal{C}_i)\leq m$ without fundamentally altering the complexity or cyclic topology of that graphical model.
\item \textit{Internal Local Constraint Removal Property:} The removal of an internal local constraint from a $q^m$-ary graphical model results in a $q^m$-ary graphical model for a new code defined on same index set.
\item \textit{Internal Local Constraint Redefinition Property:} Any internal local constraint $\mathcal{C}_i$ in a $q^m$-ary graphical model satisfying $n(\mathcal{C}_i)-k(\mathcal{C}_i)=m^\prime\leq m$ can be equivalently represented by $m^\prime$ $q$-ary single parity-check equations over the visible variable index set.
\end{enumerate} 
These properties, which are defined in detail in the appendix, are particularly useful in concert.  Specifically, let $\mathcal{G_C}$ be a $q^m$-ary graphical model for the linear code $\mathcal{C}$ over $\mathbb{F}_q$ defined on an index set $I$.  Suppose that the internal constraint $\mathcal{C}_r$ satisfying $n(\mathcal{C}_r)-k(\mathcal{C}_r)=m^\prime\leq m$ is removed from $\mathcal{G_C}$ resulting in the new code $\mathcal{C}^{\setminus r}$.  Denote by $\mathcal{C}_r^{(1)},\ldots,\mathcal{C}_r^{(m^\prime)}$ the set of $m^\prime$ $q$-ary single parity-check equations that result when $\mathcal{C}_r$ is redefined over $I$.  A vector in $\mathbb{F}_q^I$ is a codeword in $\mathcal{C}$ if and only if it is contained in $\mathcal{C}^{\setminus r}$ and satisfies each of these $m^\prime$ single parity-check equations so that
\begin{equation}
\mathcal{C}=\mathcal{C}^{\setminus r}\cap\mathcal{C}_r^{(1)}\cap\cdots\cap\mathcal{C}_r^{(m^\prime)}.
\end{equation}

\subsection{The Minimal Tree Complexity of a Code}\label{tree_comp_subsec}
The minimal trellis complexity $s\left(\mathcal{C}\right)$ of a linear code $\mathcal{C}$ over $\mathbb{F}_q$ is defined as the base-$q$ logarithm of the maximum hidden variable alphabet size in its minimal trellis \cite{LaVa95}.  Considerable attention has been paid to this quantity (cf. \cite{LaVa95,Mu88,BeBe93,KaTiFuLi93,KaTaFuLi93b,Fo94b}) as it is closely related to the important, and difficult, study of determining the minimum possible complexity of optimal SISO decoding of a given code.  This work introduces the minimal tree complexity of a linear code as a generalization of minimal trellis complexity to arbitrary cycle-free graphical model topologies.
\vspace{3pt}\begin{numDef}
The \textit{minimal tree complexity} of a linear code $\mathcal{C}$ over $\mathbb{F}_q$ is the smallest integer $t(\mathcal{C})$ such that there exists a cycle-free $q^{t(\mathcal{C})}$-ary graphical model for $\mathcal{C}$.
\end{numDef}\vspace{3pt}

Much as $s(\mathcal{C})=s(\mathcal{C}^\perp)$, the minimal tree complexity of a code $\mathcal{C}$ is equal to that of its dual.  
\begin{numProp}
Let $\mathcal{C}$ be a linear code over $\mathbb{F}_q$ with dual $\mathcal{C}^\perp$.  Then
\begin{equation}
t(\mathcal{C})=t(\mathcal{C}^\perp).
\end{equation} 
\begin{proof}
The dualizing procedure described by Forney \cite{Fo01} can be applied to a $q^{t(\mathcal{C})}$-ary graphical model for $\mathcal{C}$ in order to obtain a graphical model for $\mathcal{C}^\perp$ which is readily shown to be $q^{t(\mathcal{C})}$-ary.
\end{proof}
\end{numProp}\vspace{3pt}

\begin{figure}[htbp]
\begin{center}
\includegraphics[width=1.25in]{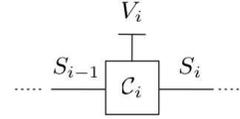}
\caption{The $q^{s(\mathcal{C})}$-ary graphical model representation of a trellis section.}
\label{trellis_sec_fig}
\end{center}
\end{figure}

Since a trellis is a cycle-free graphical model, $t\left(\mathcal{C}\right)\leq s\left(\mathcal{C}\right)$,  and all known upper bounds on $s\left(\mathcal{C}\right)$ extend to $t\left(\mathcal{C}\right)$.  Specifically, consider the section of a minimal trellis for $\mathcal{C}$ illustrated in Figure \ref{trellis_sec_fig}.  
The hidden (state) variables have alphabet sizes $|T_i|\leq s(\mathcal{C})$ and $|T_{i-1}|\leq s(\mathcal{C})$, respectively.  The local constraint $\mathcal{C}_i$ has length
\begin{equation}
n(\mathcal{C}_i)=|T_i|+|T_{i-1}|+1
\end{equation}
and dimension
\begin{equation}
k(\mathcal{C}_i)=1+\min\left(|T_i|,|T_{i-1}|\right)
\end{equation}
so that
\begin{equation}
n(\mathcal{C}_i)-k(\mathcal{C}_i)=\max\left(|T_i|,|T_{i-1}|\right)\leq s(\mathcal{C}).
\end{equation}
Lower bounds on $s\left(\mathcal{C}\right)$ do not, however, necessarily extend to $t\left(\mathcal{C}\right)$.  For example, the minimal trellis complexity of a length $2^m$, dimension $m+1$, binary first-order Reed Muller code is known to be $m$ for $m\geq 3$ \cite{BeBe93}, whereas the conditionally cycle-free generalized Tanner graphs for these codes described in \cite{HaCh06e} have a natural interpretation as $2^{m-1}$-ary cycle-free graphical models.  The Cut-Set Bound, however, precludes $t(\mathcal{C})$ from being significantly smaller than $s(\mathcal{C})$ \cite{Fo01,Fo03}.

The following lemma concerning minimal tree complexity will be used in the proof of Theorem \ref{ti_csb_theorem} in Section \ref{ticsb_sec}.
\begin{numLem}\label{tree_comp_growth_lemma}
Let $\mathcal{C}$ and $\mathcal{C}^{SPC}$ be linear codes over $\mathbb{F}_q$ defined on the index sets $I$ and $J\subseteq I$, respectively, such that $\mathcal{C}^{SPC}$ is a $q$-ary single parity-check code.  Define by $\widetilde{\mathcal{C}}$ the intersection of $\mathcal{C}$ and $\mathcal{C}^{SPC}$: 
\begin{equation}
\widetilde{\mathcal{C}}=\mathcal{C}\cap\mathcal{C}^{SPC}.
\end{equation}
The minimal tree complexity of $\widetilde{\mathcal{C}}$ is upper-bounded by
\begin{equation}
t(\widetilde{\mathcal{C}})\leq t(\mathcal{C})+1.
\end{equation}
\begin{proof}
By explicit construction of a $q^{t(\mathcal{C})+1}$-ary graphical model for $\widetilde{\mathcal{C}}$.  Let $\mathcal{G}_\mathcal{C}$ be some $q^{t(\mathcal{C})}$-ary cycle-free graphical model for $\mathcal{C}$ and let $\mathcal{T}$ be a minimal connected subtree of $\mathcal{G}_\mathcal{C}$ containing the set of $|J|$ interface constraints which involve the visible variables in $J$.  Denote by $I_{S}(\mathcal{T})\subseteq I_S$ and $I_{C}(\mathcal{T})\subseteq I_C$ the subset of hidden variables and local constraints, respectively, contained in $\mathcal{T}$.  Choose some local constraint vertex $\mathcal{C}_{\Lambda}$, $\Lambda\in I_{C}(\mathcal{T})$, as a root for $\mathcal{T}$.  Observe that the choice of $\mathcal{C}_\Lambda$, while arbitrary, induces a directionality in $\mathcal{T}$: \textit{downstream} toward the root vertex or \textit{upstream} away from the root vertex.  For every $S_i$, $i\in I_{S}(\mathcal{T})$, denote by $J_{i,\uparrow}\subseteq J$ the subset of visible variables in $J$ which are upstream from that hidden variable edge.  

A $q^{t(\mathcal{C})+1}$-ary graphical model for $\widetilde{\mathcal{C}}$ is then constructed from $\mathcal{G}_\mathcal{C}$ by updating each hidden variable $S_i$, $i\in I_{S}(\mathcal{T})$, to also contain the $q$-ary partial parity of the upstream visible variables in $J_{i,\uparrow}\subseteq J$.  The local constraints $\mathcal{C}_j$, $j\in I_{C}(\mathcal{T})\setminus\Lambda$, are updated accordingly.  Finally, $\mathcal{C}_{\Lambda}$ is updated to enforce the $q$-ary single parity constraint defined by $\mathcal{C}^{SPC}$.  This updating procedure increases the alphabet size of each hidden variable  $S_i$, $i\in I_{S}(\mathcal{T})$, by at most one and adds at most one single parity-check (or repetition) constraint to the definition of each $\mathcal{C}_j$, $j\in I_{C}(\mathcal{T})$, and the resulting cycle-free graphical model is thus at most $q^{t(\mathcal{C})+1}$-ary.
\end{proof}
\end{numLem}
\vspace{3pt}

The proof of Lemma \ref{tree_comp_growth_lemma} is detailed further by example in the appendix.
\section{The Tradeoff Between Cyclic Topology and Complexity}\label{ticsb_sec}
\subsection{The Cut-Set and Square-Root Bounds}\label{ticsb_motivation_subsec}
Wiberg's Cut-Set Bound (CSB) \cite{WiLoKo95b,Wi96} is stated below without proof in the language of Section \ref{background_sec}.  
\begin{numThm}[Cut-Set Bound]\label{csb_theorem}
Let $\mathcal{C}$ be a linear code over $\mathbb{F}_q$ defined on the index set $I$.  Let $\mathcal{G_C}$ be a graphical model for $\mathcal{C}$ containing a cut $\mathcal{X}$ corresponding to the hidden variables $S_i$, $i\in I_S(\mathcal{X})$, which partitions the index set into $J_1\subset I$ and $J_2\subset I$.  Let the base-$q$ logarithm of the midpoint hidden variable alphabet size of the minimal two-section trellis for $\mathcal{C}$ on the two-section time axis $\left\{J_1,J_2\right\}$ be $s_{\mathcal{X},\min}$.  The sum of the hidden variable alphabet sizes corresponding to the cut $\mathcal{X}$ is lower-bounded by
\begin{equation}
\sum_{i\in I_S(\mathcal{X})}\left|T_i\right|\geq s_{\mathcal{X},\min}.
\end{equation}  
\end{numThm}
\vspace{6pt}

The CSB provides insight into the tradeoff between cyclic topology and complexity in graphical models for codes and it is natural to explore its power to quantify this tradeoff.  Two questions which arise for a given linear code $\mathcal{C}$ over $\mathbb{F}_q$ in such an exploration are:
\begin{enumerate}
\item For a given complexity $m$, how many cycles must be contained in a $q^m$-ary graphical model for $\mathcal{C}$?
\item For a given number of cycles $N$, what is the smallest $m$ such that a $q^m$-ary model containing $N$ cycles for $\mathcal{C}$ can exist?
\end{enumerate} 

For a fixed cyclic topology, the CSB can be simultaneously applied to all cuts yielding a linear programming lower bound on the hidden variable alphabet sizes \cite{WiLoKo95b}.  For the special case of a single-cycle graphical model (i.e. a tail-biting trellis), this technique yields a simple solution \cite{CaFoVa99}:
\begin{numThm}[Square-Root Bound]\label{square_root_bound}
Let $\mathcal{C}$ be a linear code over $\mathbb{F}_q$ of even length and let $s_{\tmid,\min}(\mathcal{C})$ be the base-$q$ logarithm of the minimum possible hidden variable alphabet size of a conventional trellis for $\mathcal{C}$ at its midpoint over all coordinate orderings.  The base-$q$ logarithm of the minimum possible hidden variable alphabet size $s_{\TB}(\mathcal{C})$ of a tail-biting trellis for $\mathcal{C}$ is lower-bounded by
\begin{equation}
s_{\TB}(\mathcal{C})\geq s_{\tmid,\min}(\mathcal{C})/2.
\end{equation}
\end{numThm}
\vspace{6pt}

The square-root bound can thus be used to answer the questions posed above for a specific class of single-cycle graphical models.  For topologies richer than a single cycle, however, the aforementioned linear programming technique quickly becomes intractable.  Specifically, there are
\begin{equation}
2^{n(\mathcal{C})-1}-1
\end{equation}
ways to partition a size $n(\mathcal{C})$ visible variable index set into two non-empty, disjoint, subsets.  The number of cuts to be considered by the linear programming technique for a given cyclic topology thus grows exponentially with block length and a different minimal two-stage trellis must be constructed in order to bound the size of each of those cuts.
\subsection{Tree-Inducing Cuts}\label{tic_subsec}
Recall that a cut in a graph $\mathcal{G}$ is some subset of the edges $\mathcal{X}\subseteq\mathcal{E}$ the removal of which yields a disconnected graph.  A cut is thus defined without regard to the cyclic topology of the disconnected components which remain after its removal.  In order to provide a characterization of the tradeoff between cyclic topology and complexity which is more precise than that provided by the CSB alone, this work focuses on a specific type of cut  which is defined below.  Two useful properties of such cuts are established by Propositions \ref{ti_size_prop} and \ref{num_cycle_lemma}. 
\vspace{3pt}\begin{numDef}
Let $\mathcal{G}$ be a connected graph.  A \textit{tree-inducing cut} is some subset of edges $\mathcal{X}_T\subseteq \mathcal{E}$ the removal of which yields a tree with precisely two components.
\end{numDef}\vspace{3pt}
\begin{numProp}\label{ti_size_prop}
Let $\mathcal{G}=\left(\mathcal{V},\mathcal{E},\mathcal{H}\right)$ be a connected graph.  The size $X_T$ of \textit{any} tree-inducing cut $\mathcal{X}_T$ in $\mathcal{G}$ is precisely
\begin{equation}
X_T=\left|\mathcal{E}\right|-\left|\mathcal{V}\right|+2.
\end{equation}
\begin{proof}
It is well-known that a connected graph is a tree if and only if (cf. \cite{Di00})
\begin{equation}
\left|\mathcal{E}\right|=\left|\mathcal{V}\right|-1.
\end{equation}
Similarly, a graph composed of two cycle-free components satisfies
\begin{equation}\label{two_comp_tree_eq}
\left|\mathcal{E}\right|=\left|\mathcal{V}\right|-2.
\end{equation}
The result then follows from the observation that the size of a tree-inducing cut is the number of edges which must be removed in order to satisfy (\ref{two_comp_tree_eq}).
\end{proof}
\end{numProp}
\begin{numProp}\label{num_cycle_lemma}
Let $\mathcal{G}$ be a connected graph with tree-inducing cut size $X_T$.  The number of cycles $N_\mathcal{G}$ in $\mathcal{G}$ is lower-bounded by
\begin{equation}
N_\mathcal{G}\geq\binom{X_T}{2}.
\end{equation}
\begin{proof}
Let the removal of a tree-inducing cut $\mathcal{X}_T$ in the connected graph $\mathcal{G}$ yield the cycle-free components $\mathcal{G}_1$ and $\mathcal{G}_2$ and let $e_i\neq e_j\in\mathcal{X}_T$.  Since $\mathcal{G}_1$ ($\mathcal{G}_2$) is a tree, there is a unique path in $\mathcal{G}_1$ ($\mathcal{G}_2$) connecting $e_i$ and $e_j$.  There is thus a unique cycle in $\mathcal{G}$ corresponding to the edge pair $\{e_i,e_j\}$.  There are $\binom{X_T}{2}$ such distinct edge pairs which yields the lower bound.  Note that this is a lower bound because for certain graphs, there can exist cycles which contain more than two edges from a tree-inducing cut.
\end{proof}
\end{numProp}
\subsection{The Tree-Inducing Cut-Set Bound}
With tree-inducing cuts defined, the required properties of $q^m$-ary graphical models described and Lemma \ref{tree_comp_growth_lemma} established, the main result concerning the tradeoff between cyclic topology and graphical model complexity can now be stated and proved.
\begin{numThm}\label{ti_csb_theorem}
Let $\mathcal{C}$ be a linear code over $\mathbb{F}_q$ defined on the index set $I$ and suppose that $\mathcal{G}_\mathcal{C}$ is a $q^m$-ary graphical model for $\mathcal{C}$ with tree-inducing cut size $X_T$.  The minimal tree complexity of $\mathcal{C}$ is upper-bounded by
\begin{equation}
t(\mathcal{C})\leq mX_T.
\end{equation}
\begin{proof}
By induction on $X_T$.  Let $X_T=1$ and suppose that $e\in\mathcal{X}_T$ is the sole edge in some tree-inducing cut $\mathcal{X}_T$ in $\mathcal{G}_\mathcal{C}$.  Since the removal of $e$ partitions $\mathcal{G}_\mathcal{C}$ into disconnected cycle-free components, $\mathcal{G}_\mathcal{C}$ must be cycle-free and $t(\mathcal{C})\leq m$ by construction.

Now suppose that $X_T=x>1$ and let $e\in\mathcal{X}_T$ be an edge in some tree-inducing cut $\mathcal{X}_T$ in $\mathcal{G}_\mathcal{C}$.  By the first $q^m$-ary graphical model property of Section \ref{model_prop_subsec}, $e$ is incident on some internal local constraint $\mathcal{C}_i$ satisfying $n(\mathcal{C}_i)-k(\mathcal{C}_i)=m^\prime\leq m$.  Denote by $\mathcal{G}_{\mathcal{C}^{\setminus i}}$ the $q^{m}$-ary graphical model that results when $\mathcal{C}_i$ is removed from $\mathcal{G}_\mathcal{C}$, and by $\mathcal{C}^{\setminus i}$ the corresponding code over $I$.  The tree-inducing cut size of $\mathcal{G}_{\mathcal{C}^{\setminus i}}$ is at most $x-1$ since the removal of $\mathcal{C}_i$ from $\mathcal{G}_\mathcal{C}$ results in the removal a single vertex and at least two edges.  By the induction hypothesis, the minimal tree complexity of $\mathcal{C}^{\setminus i}$ is upper-bounded by
\begin{equation}
t(\mathcal{C}^{\setminus i})\leq m(x-1).
\end{equation} 

From the discussion of Section \ref{model_prop_subsec}, it is clear that $\mathcal{C}_i$ can be redefined as $m^\prime\leq m$ single parity check equations, $\mathcal{C}_{i}^{(j)}$ for $j\in[1,m^\prime]$, over $\mathbb{F}_q$ on $I$ such that
\begin{equation}
\mathcal{C}=\mathcal{C}^{\setminus i}\cap\mathcal{C}_{i}^{(1)}\cap\cdots\cap\mathcal{C}_{i}^{(m^\prime)}.
\end{equation}
It follows from Lemma \ref{tree_comp_growth_lemma} that
\begin{equation}
t(\mathcal{C})\leq t(\mathcal{C}^{\setminus i})+m^\prime\leq mx
\end{equation}
completing the proof.
\end{proof}
\end{numThm}

An immediate corollary to Theorem \ref{ti_csb_theorem} results when Proposition \ref{num_cycle_lemma} is applied in conjunction with the main result:
\begin{numCor}\label{ti_csb_cycle_cor}
Let $\mathcal{C}$ be a linear code over $\mathbb{F}_q$ with minimal tree complexity $t(\mathcal{C})$.  The number of cycles $N_m$ in any $q^m$-ary graphical model for $\mathcal{C}$ is lower-bounded by
\begin{equation}
N_m\geq \binom{\left\lfloor t(\mathcal{C})/m\right\rfloor}{2}.
\end{equation}
\end{numCor}
\vspace{6pt}
\subsection{Interpretation of the TI-CSB}\label{ti_csb_discussion_subsec}
Provided $t(\mathcal{C})$ is known or can be lower-bounded, the tree-inducing cut-set bound (TI-CSB) (and more specifically Corollary \ref{ti_csb_cycle_cor}) can be used to answer the questions posed in Section \ref{ticsb_motivation_subsec}.  The TI-CSB is further discussed below.
\subsubsection{The TI-CSB and the CSB}
On the surface, the TI-CSB and the CSB are similar in statement; however, there are three important differences between the two.  First, the  CSB does not explicitly address the complexity of the local constraints on either side of a given cut.  Forney provided a number of illustrative examples in \cite{Fo03} that stress the importance of characterizing graphical model complexity in terms of both hidden variable size and local constraint complexity.  Second, the CSB does not explicitly address the cyclic topology of the graphical model that results when the edges in a cut are removed.  The removal of a tree-inducing cut results in two cycle-free disconnected components and the size of a tree-inducing cut can thus be used to make statements about the complexity of optimal SISO decoding using variable conditioning in a cyclic graphical model (cf. \cite{Fo01,HaCh06e,WaRaHaKo97,AjHoMc98,AnHl98,HeCh01}).  Finally, and most fundamentally, the TI-CSB addresses the aforementioned intractability of applying the CSB to graphical models with rich cyclic topologies.
\subsubsection{The TI-CSB and the Square-Root Bound}
Theorem \ref{ti_csb_theorem} can be used to make a statement similar to Theorem \ref{square_root_bound} which is valid for all graphical models containing a single cycle.
\begin{numCor}\label{ti_csb_root_cor}
Let $\mathcal{C}$ be a linear code over $\mathbb{F}_q$ with minimal tree complexity $t(\mathcal{C})$ and let $m_1$ be the smallest integer such that there exists a $q^{m_1}$-ary graphical model for $\mathcal{C}$ which contains at most one cycle.  Then
\begin{equation}
m_1\geq t(\mathcal{C})/2.
\end{equation}
\end{numCor}
\vspace{8pt}

More generally, Theorem \ref{ti_csb_theorem} can be used to establish the following generalization of the square-root bound to graphical models with arbitrary cyclic topologies.
\begin{numCor}\label{ti_csb_r_root_cor}
Let $\mathcal{C}$ be a linear code over $\mathbb{F}_q$ with minimal tree complexity $t(\mathcal{C})$ and let $m_{\binom{r}{2}}$ be the smallest integer such that there exists a $q^{m_{\binom{r}{2}}}$-ary graphical model for $\mathcal{C}$ which contains at most $\binom{r}{2}$ cycles.  Then
\begin{equation}
m_{\binom{r}{2}}\geq t(\mathcal{C})/r.
\end{equation}
\end{numCor}
\vspace{8pt}

A linear interpretation of the logarithmic complexity statement of Corollary \ref{ti_csb_r_root_cor} yields the desired generalization of the square-root bound: an $r^{th}$-root complexity reduction with respect to the minimal tree complexity requires the introduction of at least $r(r-1)/2$ cycles.

There are few known examples of classical linear block codes which meet the square-root bound with equality.  Shany and Be'ery proved that many RM codes cannot meet this bound under \textit{any} bit ordering \cite{ShBe00}.  There does, however, exist a tail-biting trellis for the extended binary Golay code $\mathcal{C}_{G}$ which meets the square-root bound  with equality so that \cite{CaFoVa99}
\begin{equation}
s_{\tmid,\min}(\mathcal{C}_{G})=8\quad\mbox{and}\quad s_{\TB}(\mathcal{C}_{G})=4.  
\end{equation}
Given that this tail-biting trellis is a $2^4$-ary single cycle graphical model for $\mathcal{C}_{G}$, the minimal tree complexity of the the extended binary Golay code can be upper-bounded by Corollary \ref{ti_csb_root_cor} as
\begin{equation}
t(\mathcal{C}_{G})\leq 8.
\end{equation}
Note that the minimal bit-level conventional trellis for $\mathcal{C}_{G}$ contains (non-central) state variables with alphabet size $512$ and is thus a $2^9$-ary graphical model \cite{Mu88}.  The proof of Lemma \ref{tree_comp_growth_lemma} provides a recipe for the construction of a $2^8$-ary cycle-free graphical model for $\mathcal{C}_{G}$ from its tail-biting trellis.  It remains open as to where the minimal tree complexity of $\mathcal{C}_{G}$ is precisely $8$, however.
\subsubsection{Aymptotics of the TI-CSB}
Denote by $N_m$ the minimum number of cycles in any $q^m$-ary graphical model for a linear code $\mathcal{C}$ over $\mathbb{F}_q$ with minimal tree complexity $t(\mathcal{C})$.  For large values of $t(\mathcal{C})/m$, the lower bound on $N_m$ established by Corollary \ref{ti_csb_cycle_cor} becomes
\begin{equation}
N_m\geq \binom{\left\lfloor t(\mathcal{C})/m\right\rfloor}{2}\approx \frac{t(\mathcal{C})^2}{2m^2}.
\end{equation}   
The ratio of the minimal complexity of a cycle-free model for $\mathcal{C}$ to that of an $q^m$-ary graphical model is thus upper-bounded by
\begin{equation}
\frac{q^{t(\mathcal{C})}}{q^m}\lessapprox q^{2m\sqrt{N_m}}.
\end{equation}

\begin{figure}[ht]
\begin{center}
\includegraphics[width=3.00in]{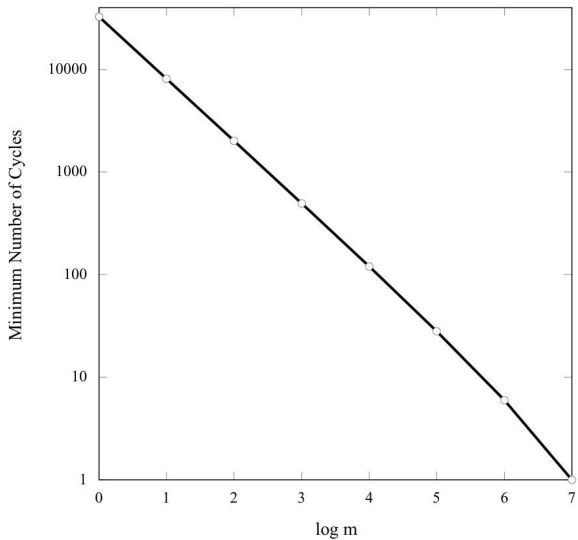}
\caption{Minimum number of cycles required for $2^m$-ary graphical models of the binary image of the $[255,223,33]$ Reed-Solomon code.}
\label{rs255_asymp_plot}
\end{center}
\end{figure}

In order to further explore the asymptotics of the tree-inducing cut-set bound, consider a code of particular practical interest: the binary image $\mathcal{C}_{RS|\mathbb{F}_2}$ of the $[255,223,33]$ Reed-Solomon code $\mathcal{C}_{RS}$.  Since $\mathcal{C}_{RS}$ is maximum distance separable, a reasonable estimate for the minimal tree complexity of this code is obtained from Wolf's bound \cite{Wo78}
\begin{equation}\label{tc_256_assume}
t(\mathcal{C}_{RS|\mathbb{F}_2})\approx 8(n(\mathcal{C}_{RS})-k(\mathcal{C}_{RS}))=256.
\end{equation}
Figure \ref{rs255_asymp_plot} plots $N_m$ as a function of $m$ for $\mathcal{C}_{RS|\mathbb{F}_2}$ assuming (\ref{tc_256_assume}).  Note that since the complexity of the decoding algorithms implied by $2^m$-ary graphical models grow roughly as $2^m$, $\log m$ is roughly a $\log\log$ decoding complexity measure. 
\subsubsection{On Complexity Measures}
Much as there are many valid complexity measures for conventional trellises, there are many reasonable metrics for the measurement of cyclic graphical model complexity.  While there exists a unique minimal trellis for any linear block code which simultaneously minimizes all reasonable measures of complexity \cite{Mc96}, even for the class cyclic graphical models with the most basic cyclic topology - tail-biting trellises - minimal models are not unique \cite{KoVa03c}.  The complexity measure introduced by this work was motivated by the desire to have a metric which simultaneously captures hidden variable complexity and local constraint complexity thus disallowing local constraints from ``hiding" complexity.  There are many conceivable measures of local constraint complexity: one could upper-bound the state complexity of the local constraints or even their minimal tree complexity (thus defining minimal tree complexity recursively).  The local constraint complexity measure used in this work is essentially Wolf's bound \cite{Wo78} and is thus a potentially conservative \textit{upper bound} on any reasonable measure of local constraint decoding complexity.
\section{Graphical Model Transformation}\label{model_tx_sec}
Let $\mathcal{G_C}$ be a graphical model for the linear code $\mathcal{C}$ over $\mathbb{F}_q$.  This work introduces eight \textit{basic graphical model operations} the application of which to $\mathcal{G_C}$ results in a new graphical model for $\mathcal{C}$:
\begin{itemize}
\item The merging of two local constraints $\mathcal{C}_{i_1}$ and $\mathcal{C}_{i_2}$ into the new local constraint $\mathcal{C}_i$ which satisfies
\begin{equation}
\mathcal{C}_i=\mathcal{C}_{i_1}\cap\;\mathcal{C}_{i_2}.
\end{equation}
\item The splitting of a local constraint $\mathcal{C}_j$ into two new local constraints $\mathcal{C}_{j_1}$ and $\mathcal{C}_{j_2}$ which satisfy
\begin{equation}
\mathcal{C}_{j_1}\cap\;\mathcal{C}_{j_2}=\mathcal{C}_j.
\end{equation}
\item The insertion/removal of a degree-$2$ repetition constraint.
\item The insertion/removal of a trival length $0$, dimension $0$ local constraint.
\item The insertion/removal of an isolated partial parity-check constraint.
\end{itemize}
Note that some of these operations have been introduced implicitly in this work and others already.  For example, the proof of the local constraint involvement property of $q^m$-ary graphical models presented in Section \ref{model_prop_subsec} utilizes degree-$2$ repetition constraint insertion.  Local constraint merging has been considered by a number of authors under the rubric of clustering (e.g. \cite{KsFrLo01,Fo01}).  This work introduces the term merging specifically so that it can be contrasted with its inverse operation: splitting.  Detailed definitions of each of the eight basic graphical model operations  are given in the appendix.  In this section, it is shown that these basic operations span the entire space of graphical models for $\mathcal{C}$.
\begin{numThm}\label{basic_move_thm}
Let $\mathcal{G_C}$ and $\widetilde{\mathcal{G}}_{\mathcal{C}}$ be two graphical models for the linear code $\mathcal{C}$ over $\mathbb{F}_q$.  Then $\mathcal{G_C}$ can be transformed into $\widetilde{\mathcal{G}}_{\mathcal{C}}$ via the application of a \textit{finite} number of basic graphical model operations.

\begin{proof}
Define the following four sub-transformations which can be used to transform $\mathcal{G}_{\mathcal{C}}$ into a Tanner graph $\mathcal{G}_{\mathcal{C}}^{T}$:
\begin{enumerate}
\item The transformation of $\mathcal{G}_{\mathcal{C}}$ into a $q$-ary model $\mathcal{G}_{\mathcal{C}}^{q}$.
\item The transformation of $\mathcal{G}_{\mathcal{C}}^{q}$ into a (possibly) redundant generalized Tanner graph $\mathcal{G}_{\mathcal{C}}^{r}$.
\item The transformation of $\mathcal{G}_{\mathcal{C}}^{r}$ into a non-redundant generalized Tanner graph $\mathcal{G}_{\mathcal{C}}^{g}$.
\item The transformation of $\mathcal{G}_{\mathcal{C}}^{g}$ into a Tanner graph $\mathcal{G}_{\mathcal{C}}^{T}$.
\end{enumerate}
Since each basic graphical model operation has an inverse, $\mathcal{G}_{\mathcal{C}}^{T}$ can be transformed into $\mathcal{G}_{\mathcal{C}}$ by inverting each of the four sub-transformations.  In order to prove that $\mathcal{G}_{\mathcal{C}}$ can be transformed into $\widetilde{\mathcal{G}}_{\mathcal{C}}$ via the application of a finite number of basic graphical model operations, it suffices to show that each of the four sub-transformations requires a finite number of operations and that the transformation of the Tanner graph $\mathcal{G}_{\mathcal{C}}^{T}$ into a Tanner graph $\widetilde{\mathcal{G}}_{\mathcal{C}}^{T}$ corresponding to $\widetilde{\mathcal{G}}_{\mathcal{C}}$ requires a finite number of operations.  This proof summary is illustrated in Figure \ref{model_tx_mech_fig}.
\begin{figure}[h] 
\begin{center}
\includegraphics[width=1.55in]{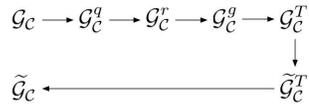}
\caption{The transformation of $\mathcal{G}_{\mathcal{C}}$ into $\widetilde{\mathcal{G}}_{\mathcal{C}}$ via five sub-transformations.}
\label{model_tx_mech_fig}
\end{center}
\end{figure}  

That each of the five sub-transformations from $\mathcal{G}_{\mathcal{C}}$ to $\widetilde{\mathcal{G}}_{\mathcal{C}}^T$ illustrated in Figure \ref{model_tx_mech_fig} requires only a finite number of basic graphical model operations is proved below.
\setcounter{subsubsection}{0}
\subsubsection{$\mathcal{G}_{\mathcal{C}}\rightarrow\mathcal{G}_{\mathcal{C}}^{q}$}
The graphical model $\mathcal{G}_{\mathcal{C}}$ is transformed into the $q$-ary model $\mathcal{G}_{\mathcal{C}}^{q}$ as follows.  Each local constraint $\mathcal{C}_i$ in $\mathcal{G}_{\mathcal{C}}$ is split into the $n(\mathcal{C}_i)-k(\mathcal{C}_i)$ $q$-ary single parity-check constraints which define it.    A degree-$2$ repetition constraint is then inserted into every hidden variable with alphabet index set size $m>1$ and these repetition constraints are then each split into $m$ $	q$-ary repetition constraints as illustrated in Figure \ref{model_tx_qary_fig}.  Each local constraint $\mathcal{C}_j$ in the resulting graphical model satisfies $n(\mathcal{C}_j)-k(\mathcal{C}_j)=1$.  Similarly, each hidden variable $S_j$ in the resulting graphical model satisfies $|T_j|=1$.
\begin{figure}[htbp]
\begin{center}
\includegraphics[width=3.25in]{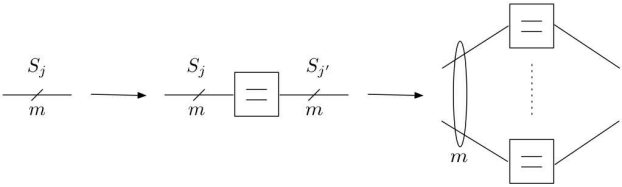}
\caption{Transformation of the $q^m$-ary hidden variable $S_j$ into $q$-ary hidden variables.}
\label{model_tx_qary_fig}
\end{center}
\end{figure}
\subsubsection{$\mathcal{G}_{\mathcal{C}}^{q}\rightarrow\mathcal{G}_{\mathcal{C}}^{r}$}A (possibly redundant) generalized Tanner graph is simply a bipartite $q$-ary graphical model with one vertex class corresponding to repetition constraints and one to single parity-check constraints in which visible variables are incident only on repetition constraints.  By appropriately inserting degree-$2$ repetition constraints, the $q$-ary model $\mathcal{G}_{\mathcal{C}}^{q}$ can be transformed into $\mathcal{G}_{\mathcal{C}}^{r}$.
\subsubsection{$\mathcal{G}_{\mathcal{C}}^{r}\rightarrow\mathcal{G}_{\mathcal{C}}^{g}$}Let the generalized Tanner graph $\mathcal{G}_{\mathcal{C}}^{r}$ correspond to an $r_H\times n(\mathcal{C})+g$ redundant parity-check matrix $H^{(r,g)}_{\mathcal{C}}$ for a degree-$g$ generalized extension of $\mathcal{C}$ with rank
\begin{equation}
\rank(H^{(r,g)}_{\mathcal{C}})=n(\mathcal{C})-k(\mathcal{C})+g.
\end{equation}
A finite number of row operations can be applied to $H^{(r,g)}_{\mathcal{C}}$ resulting in a new parity-check matrix the last $r_H-\rank(H^{(r,g)}_{\mathcal{C}})$ rows of which are all zero.  Similarly, a finite number of basic operations can be applied to $\mathcal{G}_{\mathcal{C}}^{r}$ resulting in a generalized Tanner graph containing $r_H-\rank(H^{(r,g)}_{\mathcal{C}})$ trivial constraints which can then be removed to yield $\mathcal{G}_{\mathcal{C}}^{g}$.  Specifically, consider the row operation on $H^{(r,g)}_{\mathcal{C}}$ which replaces a row $\pmb{h}_i$ by
\begin{equation}
\widetilde{\pmb{h}}_i=\pmb{h}_i+\beta_j\pmb{h}_j
\end{equation}
where $\beta_j\in\mathbb{F}_q$.  The graphical model transformation corresponding to this row operation first merges the $q$-ary single parity-check constraints $\mathcal{C}_i$ and $\mathcal{C}_j$ (which correspond to rows $\pmb{h}_i$ and $\pmb{h}_j$, respectively) and then splits the resulting check into the constraints $\widetilde{\mathcal{C}_i}$ and $\mathcal{C}_j$ (which correspond to rows $\widetilde{\pmb{h}}_i$ and $\pmb{h}_j$, respectively).  Note that this procedure is valid since
\begin{equation}
\mathcal{C}_i\cap\mathcal{C}_j=\widetilde{\mathcal{C}}_i\cap\mathcal{C}_j.
\end{equation}
\subsubsection{$\mathcal{G}_{\mathcal{C}}^{g}\rightarrow\mathcal{G}_{\mathcal{C}}^{T}$}Let the degree-$g$ generalized Tanner graph $\mathcal{G}_{\mathcal{C}}^g$ correspond to an $n(\mathcal{C})-k(\mathcal{C})+g\times n(\mathcal{C})+g$ parity-check matrix $H_\mathcal{C}^{(g)}$.  A degree-$(g-1)$ generalized Tanner graph $\mathcal{G}_{\mathcal{C}}^{g-1}$ is obtained from $\mathcal{G}_{\mathcal{C}}^{g}$ as follows.  Denote by $\widehat{H}_\mathcal{C}^{(g)}$ the  parity-check matrix for the degree-$g$ generalized extension defined by $H_\mathcal{C}^{(g)}$ which is systematic in the position corresponding to the $g$-th partial parity symbol.  Since a finite number of row operations can be applied to $H_\mathcal{C}^{(g)}$ to yield $\widehat{H}_\mathcal{C}^{(g)}$, a finite number of local constraint merge and split operations can be be applied to $\mathcal{G}_{\mathcal{C}}^{g}$ to yield the corresponding generalized Tanner graph $\widehat{\mathcal{G}}_{\mathcal{C}}^{g}$.  Removing the now isolated partial-parity check constraint corresponding to the $g$-th partial parity symbol in $\widehat{\mathcal{G}}_{\mathcal{C}}^{g}$ yields the desired degree-$(g-1)$ generalized Tanner graph $\mathcal{G}_{\mathcal{C}}^{g-1}$.  By repeatedly applying this procedure, all partial parity symbols can be removed from $\mathcal{G}_{\mathcal{C}}^{g}$ resulting in $\mathcal{G}_{\mathcal{C}}^{T}$.
\subsubsection{$\mathcal{G}_{\mathcal{C}}^{T}\rightarrow\widetilde{\mathcal{G}}_{\mathcal{C}}^{T}$}Let the Tanner graphs $\mathcal{G}_{\mathcal{C}}^{T}$ and $\widetilde{\mathcal{G}}_{\mathcal{C}}^{T}$ correspond to the parity-check matrices $H_\mathcal{C}$ and $\widetilde{H}_{\mathcal{C}}$, respectively.  Since $H_\mathcal{C}$ can be transformed into $\widetilde{H}_{\mathcal{C}}$ via a finite number of row operations, $\mathcal{G}_{\mathcal{C}}^{T}$  can be similarly transformed into $\widetilde{\mathcal{G}}_{\mathcal{C}}^{T}$ via the application of a finite number of local constraint merge and split operations.
\end{proof}
\end{numThm}\vspace{3pt}
\section{Graphical Model Extraction via Transformation}\label{extraction_sec}
The set of basic model operations introduced in the previous section enables the space of all graphical models for a given code $\mathcal{C}$ to be searched, thus allowing for model extraction to be expressed as an optimization problem.  The challenges of defining extraction as optimization are twofold.  First, a cost measure on the space of graphical models must be found which is simultaneously meaningful in some real sense (e.g. highly correlated with decoding performance) and computationally tractable.  Second, given that discrete optimization problems are in general very hard, heuristics for extraction must be found.  In this section, heuristics are investigated for the extraction of graphical models for binary linear block codes from an initial Tanner graph.  The cost measures considered are functions of the short cycle structure of graphical models.  The use of such cost measures is motivated first by empirical evidence concerning the detrimental effect of short cycles on decoding performance  (cf. \cite{Wi96,Fo01,Ta81,MaBa01,ArElHu01,GeEpSm01,TiJoViWe04,HaCh06}) and second by the existence of an efficient algorithm for counting short cycles in bipartite graphs \cite{HaCh06}.  Simulation results for the models extracted via these heuristics for a number of extended BCH codes are presented and discussed in Section \ref{sim_results_subsec}.
\vspace{-6pt}
\subsection{A Greedy Heuristic for Tanner Graph Extraction}
The Tanner graphs corresponding to many linear block codes of practical interest \textit{necessarily} contain many short cycles \cite{HaGrCh06}.  Suppose that any Tanner graph for a given code $\mathcal{C}$ must have girth at least $g_{\min}(\mathcal{C})$; an interesting problem is the extraction of a Tanner graph for $\mathcal{C}$ containing the smallest number of $g_{\min}(\mathcal{C})$-cycles.  The extraction of such Tanner graphs is especially useful in the context of ad-hoc decoding algorithms which utilize Tanner graphs such as Jiang and Narayanan's stochastic shifting based iterative decoding algorithm for cyclic codes \cite{JiNa04} and the random redundant iterative decoding algorithm presented in \cite{HaCh06c}.

Algorithm \ref{tanner_graph_alg} performs a greedy search for a Tanner graph for $\mathcal{C}$ with girth $g_{\min}(\mathcal{C})$ and the smallest number of $g_{\min}(\mathcal{C})$-cycles starting with an initial Tanner graph $\TG(H_\mathcal{C})$ which corresponds to some binary parity-check matrix $H_\mathcal{C}$.  Define an $(i,j)$-row operation as the replacement of row $\pmb{h}_j$ in $H_\mathcal{C}$ by the binary sum of rows $\pmb{h}_i$ and $\pmb{h}_j$.  As detailed in the proof of Theorem \ref{basic_move_thm}, if $\mathcal{C}_i$ and $\mathcal{C}_j$ are the single parity-check constraints in $\TG(H_\mathcal{C})$ corresponding to $\pmb{h}_i$ and $\pmb{h}_j$, respectively, then an $(i,j)$-row operation in $H_\mathcal{C}$ is equivalent to merging $\mathcal{C}_i$ and $\mathcal{C}_j$ to form a new constraint $\mathcal{C}_{i,j}=\mathcal{C}_i\cap\mathcal{C}_j$ and then splitting $\mathcal{C}_{i,j}$ into $\mathcal{C}_i$ and $\widetilde{\mathcal{C}}_j$ (where $\widetilde{\mathcal{C}}_j$ enforces the binary sum of rows $\pmb{h}_i$ and $\pmb{h}_j$).  Algorithm \ref{tanner_graph_alg} iteratively finds the rows $\pmb{h}_i$ and $\pmb{h}_j$ in $H_\mathcal{C}$ with corresponding $(i,j)$-row operation that results in the largest short cycle reduction in $\TG(H_\mathcal{C})$ at every step.  This greedy search continues until there are no more row operations that improve the short cycle structure of $\TG(H_\mathcal{C})$.  
\begin{algorithm}[htbp]
\SetLine
\KwIn{$r_H\times n(\mathcal{C})$ binary parity-check matrix $H_\mathcal{C}$.}
\KwOut{$r_H \times n(\mathcal{C})$ binary parity-check matrix $H_\mathcal{C}^\prime$.}
\BlankLine
$H_\mathcal{C}^\prime\leftarrow H_\mathcal{C}$; $i^\star\leftarrow -1$; $j^\star\leftarrow -1$; $g^\star\leftarrow $ girth of $\TG\left(H^\prime_\mathcal{C}\right)$\;
$N_{g^\star}^\star\leftarrow $ number of $g^\star$-cycles in $\TG\left(H_\mathcal{C}^\prime\right)$\;
$N_{g^\star+2}^\star\leftarrow $ number of $g^\star+2$-cycles in $\TG\left(H_\mathcal{C}^\prime\right)$\;
\Repeat{$i^\star=-1$ \& $j^\star=-1$}{
\lIf{$i^\star \neq j^\star$}{
Replace row $\pmb{h}_{j^\star}$ in $H_\mathcal{C}^\prime$ with binary sum of rows $\pmb{h}_{i^\star}$ and $\pmb{h}_{j^\star}$\;
}
$i^\star\leftarrow -1$; $j^\star\leftarrow -1$\;
\For{$i,j=0,\ldots,r_H-1$, $i\neq j$}{
Replace row $\pmb{h}_j$ in $H_\mathcal{C}^\prime$ with binary sum of rows $\pmb{h}_i$ and $\pmb{h}_j$\;
$g\leftarrow $ girth of $\TG\left(H_\mathcal{C}^\prime\right)$\;
$N_g\leftarrow $ number of $g$-cycles in $\TG\left(H_\mathcal{C}^\prime\right)$\;
$N_{g+2}\leftarrow $ number of $g+2$-cycles in $\TG\left(H_\mathcal{C}^\prime\right)$\;
\If{$g>g^\star$}{$g^\star\leftarrow g$; $i^\star\leftarrow i$; $j^\star\leftarrow j$; $N_g^\star\leftarrow N_g$; $N_{g+2}^\star\leftarrow N_{g+2}$\;}
\lElseIf{$g=g^\star$ AND $N_g<N_g^\star$}{
$i^\star\leftarrow i$; $j^\star\leftarrow j$; $N_g^\star\leftarrow N_g$; $N_{g+2}^\star\leftarrow N_{g+2}$\;}
\ElseIf{$g=g^\star$ AND $N_g=N_g^\star$}{
\lIf{$N_{g+2}<N_{g+2}^\star$}{
$i^\star\leftarrow i$; $j^\star\leftarrow j$; $N_{g+2}^\star\leftarrow N_{g+2}$\;
}
}
Undo row replacement\;
}
}
\BlankLine
\Return{$H_\mathcal{C}^\prime$}

\
\caption{Greedy heuristic for the reduction of short cycles in Tanner graphs for binary codes.}\label{tanner_graph_alg}\end{algorithm}
\vspace{-6pt}
\subsection{A Greedy Heuristic for Generalized Tanner Graph Extraction}
A number of authors have studied the extraction of generalized Tanner graphs (GTGs) of codes for which $g_{\min}(\mathcal{C})=4$ with a particular focus on models which are $4$-cycle-free and which correspond to generalized code extensions of minimal degree \cite{SaVa05,KuMi05}.  Minimal degree extensions are sought because no information is available to the decoder about the partial parity symbols in a generalized Tanner graph and the introduction of too many such symbols has been observed empirically to adversely affect decoding performance \cite{KuMi05}.

Generalized Tanner graph extraction algorithms proceed via the insertion of partial parity symbols, an operation which is most readily described as a parity-check matrix manipulation\footnote{Note that partial parity insertion can also be viewed through the lens of graphical model transformation.  The insertion of partial parity symbol proceeds via the insertion of an isolated partial parity check followed by a series of local constraint merge and split operations.}.  Following the notation introduced in Section \ref{tg_gtg_subsec}, suppose that a partial parity on the coordinates indexed by
\begin{equation}
J\subseteq I\cup\left\{p_1,p_2,\ldots,p_g\right\}
\end{equation}  
is to be introduced to a GTG for $\mathcal{C}$ corresponding to a degree-$g$ generalized extension $\widetilde{\mathcal{C}}$ with parity-check matrix $H_{\widetilde{\mathcal{C}}}$.  A row $\pmb{h}_p$ is first appended to $H_{\widetilde{\mathcal{C}}}$ with a $1$ in the positions corresponding to coordinates indexed by $J$ and a $0$ in the other positions.  A column is then appended to $H_{\widetilde{\mathcal{C}}}$ with a $1$ only in the position corresponding to $\pmb{h}_p$.  The resulting parity-check matrix $H_{\widehat{\mathcal{C}}}$ describes a degree-$g+1$ generalized extension $\widehat{\mathcal{C}}$.  Every row $\pmb{h}_i\neq \pmb{h}_p$ in $H_{\widehat{\mathcal{C}}}$ which contains a $1$ in all of the positions corresponding to coordinates  indexed by $J$ is then replaced by the binary sum of $\pmb{h}_i$ and $\pmb{h}_p$.  Suppose that there are $r(J)$ such rows.  It is readily verified that the tree-inducing cut size $\widehat{X}_T$ of the GTG that results from this insertion is related to that of the initial GTG, $\widetilde{X}_T$, by
\begin{equation}\label{gtg_xt_red_eq}
\Delta X_T=\widetilde{X}_T-\widehat{X}_T=(|J|-1)(r(J)-1).
\end{equation}

Algorithm \ref{gtg_alg} performs a greedy search for a $4$-cycle-free generalized Tanner graph for $\mathcal{C}$ with the smallest number of inserted partial parity symbols starting with an initial Tanner graph $\TG(H_\mathcal{C})$ which corresponds to some binary parity-check matrix $H_\mathcal{C}$.  Algorithm \ref{gtg_alg} iteratively finds the symbol subsets that result in the largest tree-inducing cut size reduction and then introduces the partial parity symbol corresponding to one of those subsets.  At each step, Algorithm \ref{gtg_alg} uses Algorithm \ref{gtg_subalg} to generate a candidate list of partial parity symbols to insert and chooses from that list the symbol which reduces the most short cycles when inserted.  This greedy procedure continues until the generalized Tanner graph contains no $4$-cycles.   

\begin{figure}[h]
\begin{center}
\includegraphics[width=3.00in]{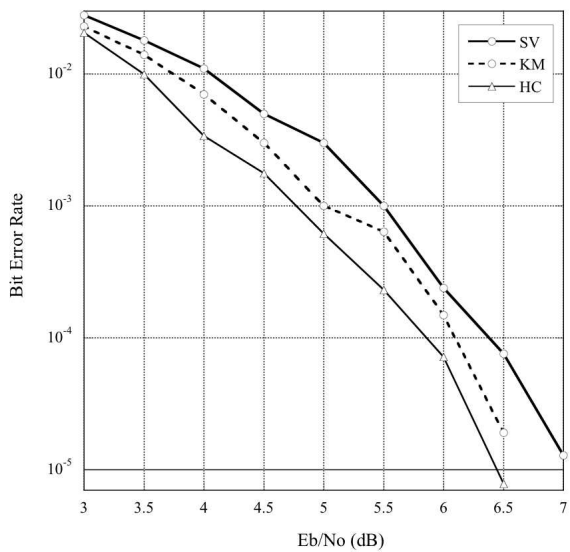}
\caption{Bit error rate performance of three GTG decoding algorithms for the $[31,21,5]$ BCH code. One-hundred iterations of a flooding schedule were performed.  Binary antipodal signaling over an AWGN channel is assumed.}
\label{bch_31_21_fig}
\end{center}
\end{figure}

Algorithm \ref{gtg_alg} is closely related to the GTG extraction heuristics proposed by Sankaranarayanan and Vasi\'{c} in \cite{SaVa05} and Kumar and Milenkovic in \cite{KuMi05} (henceforth referred to as the SV and KM heuristics, respectively).  It is readily shown that Algorithm \ref{gtg_alg} is guaranteed to terminate using the proof technique of \cite{SaVa05}. The SV heuristic considers only the insertion of partial parity symbols corresponding to coordinate index sets of size 2 (i.e. $|J|=2$).  The KM heuristic considers only the insertion of partial parity symbols corresponding to coordinate index sets satisfying $r(J)=2$.  Algorithm \ref{gtg_subalg}, however, considers all coordinate index sets satisfying $|J|=2,3,4$ and $r(J)=2,3,4$ and then uses (\ref{gtg_xt_red_eq}) to evaluate which of these coordinate sets results in the largest tree-inducing cut size reduction.  Algorithm \ref{gtg_alg} is thus able to extract GTGs corresponding to generalized extensions of smaller degree than the SV and KM heuristics.  In order to illustrate this observation, the degrees of the generalized code extensions that result when the SV, KM and proposed (HC) heuristics are applied to parity-check matrices for three codes are provided in Table \ref{gtg_table}.  Figure \ref{bch_31_21_fig} compares the performance of the three extracted GTG decoding algorithms for the $[31,21,5]$ BCH code in order to illustrate the efficacy of extracting GTGs corresponding to extensions of smallest possible degree.
\begin{table}[h]
\begin{center}
\begin{tabular}{c||c|c|c}
Code&SV&KM&HC\\
\hline\hline
$[23,12,7]$ Golay&$18$&$11$&$10$\\
\hline
$[31,21,5]$ BCH&$47$&$19$&$12$\\
\hline
$[63,30,13]$ BCH&$264$&$121$&$69$
\end{tabular}
\caption{Generalized code extension degrees corresponding to the $4$-cycle-free GTGs obtained via the SV, KM, and HC heuristics.}\label{gtg_table}
\end{center}
\end{table}
\vspace{-6pt}
\begin{algorithm}[htbp]
\SetLine
\KwIn{Binary generalized parity-check matrix $H_{\widetilde{\mathcal{C}}}$.}
\KwOut{List $\mathcal{S}$ of best partial parity symbols sets.}
\BlankLine
$\mathcal{S}\leftarrow\emptyset$; $\Delta X_{T}^\star\leftarrow 0$\;
\BlankLine
\tcp{Consider pairs of columns of $H_{\widetilde{\mathcal{C}}}$.}
$J_2\leftarrow$ coordinate pair that maximizes $r(J_2)$\;
Append $J_2$ to $\mathcal{S}$; $\Delta X_{T}^\star\leftarrow r(J_2)-1$\;
\BlankLine
\tcp{Consider $3$-tuples of columns of $H_{\widetilde{\mathcal{C}}}$.}
$J_3\leftarrow$ coordinate $3$-tuple that maximizes $r(J_3)$\;
\If{$2(r(J_3)-1)>\Delta X_{T}^\star$}{$\mathcal{S}\leftarrow\emptyset$; Append $J_3$ to $\mathcal{S}$; $\Delta X_{T}^\star\leftarrow 2(r(J_3)-1)$\;}
\lElseIf{$2(r(J_3)-1)=\Delta X_{T}^\star$}{Append $J_3$ to $\mathcal{S}$\;}
\BlankLine
\tcp{Consider $4$-tuples of columns of $H_{\widetilde{\mathcal{C}}}$.}
$J_4\leftarrow$ coordinate $4$-tuple that maximizes $r(J_4)$\;
\If{$3(r(J_4)-1)>\Delta X_{T}^\star$}{$\mathcal{S}\leftarrow\emptyset$; Append $J_4$ to $\mathcal{S}$; $\Delta X_{T}^\star\leftarrow 3(r(J_4)-1)$\;}
\lElseIf{$3(r(J_4)-1)=\Delta X_{T}^\star$}{Append $J_4$ to $\mathcal{S}$\;}
\BlankLine
\tcp{Consider pairs of rows of $H_{\widetilde{\mathcal{C}}}$.}
$J_i\leftarrow$ largest coordinate subset such that $r(J_i)=2$\;
\If{$|J_i|-1>\Delta X_T^\star$}{$\mathcal{S}\leftarrow\emptyset$; Append $J_i$ to $\mathcal{S}$; $\Delta X_{T}^\star\leftarrow |J_i|-1$\;}
\lElseIf{$|J_i|-1=\Delta X_T^\star$}{Append $J_i$ to $\mathcal{S}$\;}
\BlankLine
\tcp{Consider $3$-tuples of rows of $H_{\widetilde{\mathcal{C}}}$.}
$J_j\leftarrow$ largest coordinate subset such that $r(J_j)=3$\;
\If{$2(|J_j|-1)>\Delta X_T^\star$}{$\mathcal{S}\leftarrow\emptyset$; Append $J_j$ to $\mathcal{S}$; $\Delta X_{T}^\star\leftarrow 2(|J_j|-1)$\;}
\lElseIf{$2(|J_j|-1)=\Delta X_T^\star$}{Append $J_j$ to $\mathcal{S}$\;}
\BlankLine
\tcp{Consider $4$-tuples of rows of $H_{\widetilde{\mathcal{C}}}$.}
$J_k\leftarrow$ largest coordinate subset such that $r(J_k)=4$\;
\If{$3(|J_k|-1)>\Delta X_T^\star$}{$\mathcal{S}\leftarrow\emptyset$; Append $J_k$ to $\mathcal{S}$; $\Delta X_{T}^\star\leftarrow 3(|J_k|-1)$\;}
\lElseIf{$3(|J_k|-1)=\Delta X_T^\star$}{Append $J_k$ to $\mathcal{S}$\;}
\BlankLine
\Return{$\mathcal{S}$}

\
\caption{Heuristic for generating candidate partial parity symbols.}\label{gtg_subalg}\end{algorithm}

\begin{algorithm}[htbp]
\SetLine
\KwIn{Binary parity-check matrix $H_\mathcal{C}$.}
\KwOut{Binary generalized parity-check matrix $H_{\widetilde{\mathcal{C}}}$.}
\BlankLine
$H_{\widetilde{\mathcal{C}}}\leftarrow H_\mathcal{C}$\;
\While{$\GTG(H_{\widetilde{\mathcal{C}}})$ contains $4$-cycles}{
$\mathcal{S}\leftarrow$ set of candidate partial parity symbol
 
\
$\phantom{\mathcal{S}\leftarrow}$subsets from Algorithm \ref{gtg_subalg}\;
$J^\star\leftarrow$ subset in $\mathcal{S}$ the insertion of which reduces

\
$\phantom{J^\star\leftarrow}$the most $4$--cycles in $\GTG(H_{\widetilde{\mathcal{C}}})$\;
Insert  symbol corresponding to $J^\star$ in $H_{\widetilde{\mathcal{C}}}$\;
}
\Return{$H_{\widetilde{\mathcal{C}}}$}

\
\caption{Greedy heuristic for the removal of $4$-cycles in binary generalized Tanner graphs.}\label{gtg_alg}\end{algorithm}
\newpage
\subsection{A Greedy Heuristic for $2^m$-ary Model Extraction}
For most codes, the decoding algorithms implied by generalized Tanner graphs exhibit only modest gains with respect to those implied by Tanner graphs, if any, thus motivating the search for more complex graphical models.  Algorithm \ref{twom_alg} iteratively applies the constraint merging operation in order to obtain a $2^{m^\star}$-ary graphical model from an initial Tanner graph $\TG(H_\mathcal{C})$ for some prescribed maximum complexity $m^\star$.  At each step, Algorithm \ref{twom_alg} determines the pair of local constraints $\mathcal{C}_i$ and $\mathcal{C}_j$ which when merged reduces the most short cycles without violating the maximum complexity constraint $m^\star$.  In order to ensure that that the efficient cycle counting algorithm of \cite{HaCh06} can be utilized, only pairs of checks which are both internal or both interface are merged at each step.  Since the initial Tanner graph is bipartite with vertex classes corresponding to interface (repetition) and internal (single parity-check) constraints, the graphical models that result from every such local constraint merge operations are similarly bipartite.

\begin{algorithm}[htbp]
\SetLine
\KwIn{Tanner graph $\TG(H_\mathcal{C})$.  Max. complexity $m^\star$.}
\KwOut{$2^{m^\star}$-ary graphical model $\GM$ for $\mathcal{C}$.}
\BlankLine
$\GM\leftarrow\TG(H_\mathcal{C})$\;
\Repeat{No allowed $4$-cycle reducing merge operations remain}
{$(\mathcal{C}_i,\mathcal{C}_j)\leftarrow$ pair of incident or internal constraints

\
$\phantom{(\mathcal{C}_i,\mathcal{C}_j)\leftarrow}$the removal of which removes the most

\
$\phantom{(\mathcal{C}_i,\mathcal{C}_j)\leftarrow}$$4$-cycles from $\GM$ while not violating

\
$\phantom{(\mathcal{C}_i,\mathcal{C}_j)\leftarrow}$the $2^{m^\star}$-ary complexity constraint\;
Merge local constraints $\mathcal{C}_i$ and $\mathcal{C}_j$ in $\GM$\;
}
\Return{$\GM$}

\
\caption{Greedy heuristic for the extraction of $2^m$-ary graphical models.}\label{twom_alg}\end{algorithm}
\vspace{-18pt}
\subsection{Simulation Results}\label{sim_results_subsec}
\begin{figure}[t]
\begin{center}
\includegraphics[width=3.0in]{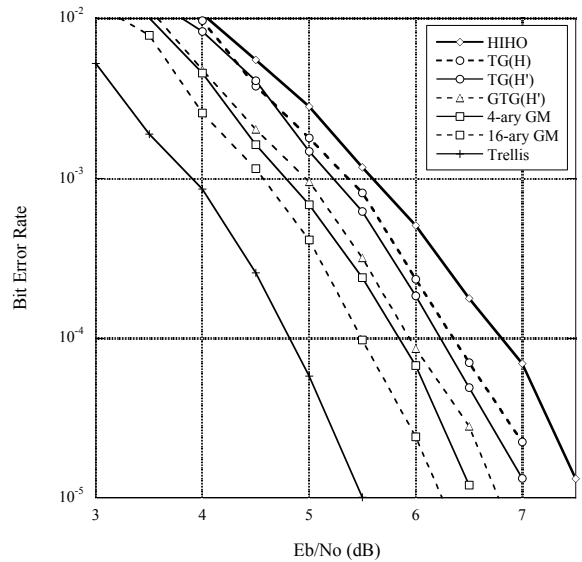}
\caption{Bit error rate performance of different decoding algorithms for the $[32,21,6]$ extended BCH code.  Fifty iterations of a flooding schedule were performed for all of the suboptimal SISO decoding algorithms.}
\label{bch_32_21_fig}
\end{center}
\end{figure}

\begin{figure}[t]
\begin{center}
\includegraphics[width=3.0in]{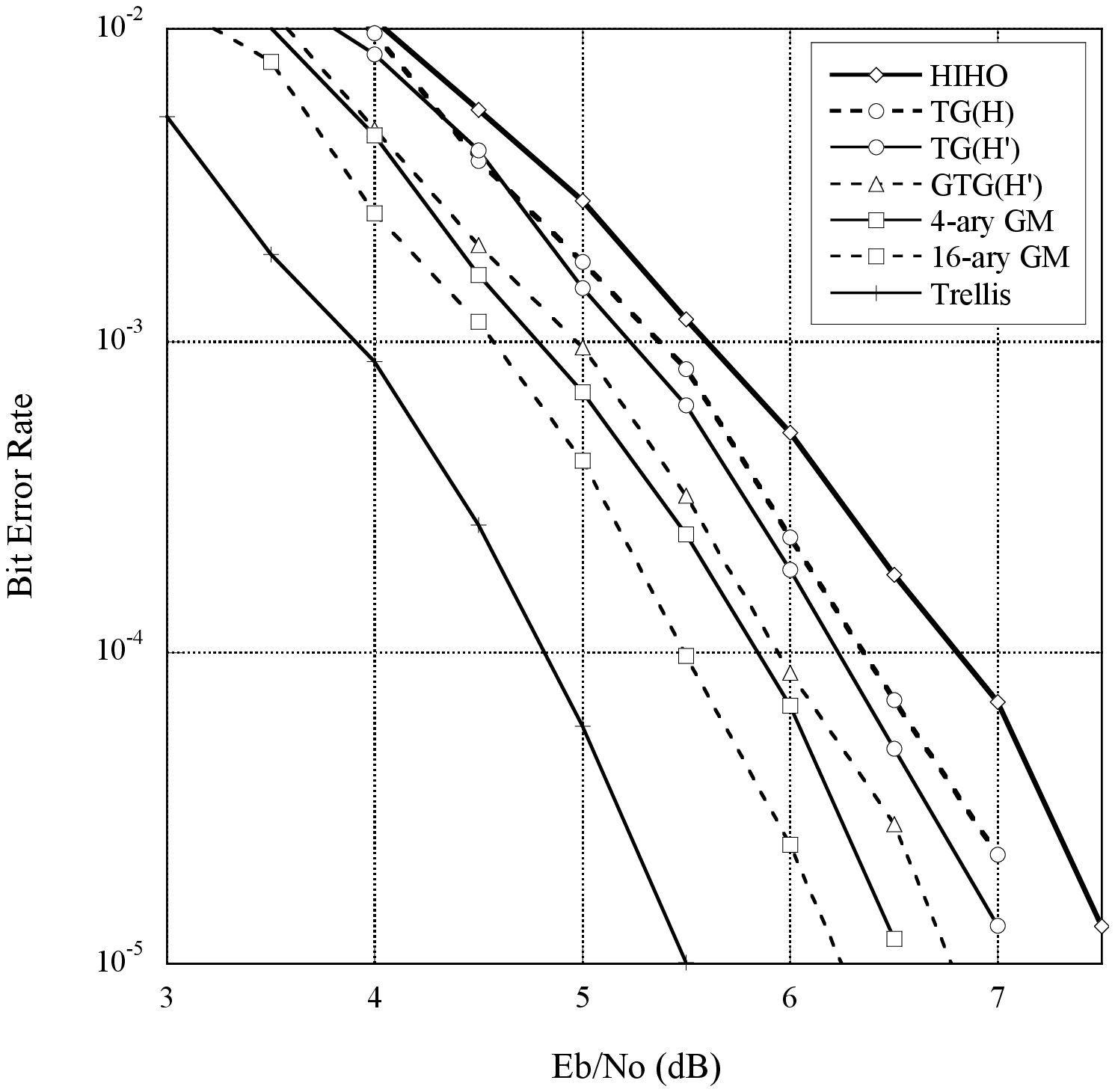}
\caption{Bit error rate performance of different decoding algorithms for the $[64,51,6]$ extended BCH code.  Fifty iterations of a flooding schedule were performed for all of the suboptimal SISO decoding algorithms.}
\label{bch_64_51_fig}
\end{center}
\end{figure}

The proposed extraction heuristics were applied to two extended BCH codes with parameters $[32,21,6]$ and $[64,51,6]$, respectively.  In both Figures \ref{bch_32_21_fig} and \ref{bch_64_51_fig} the performance of a number of suboptimal SISO decoding algorithms for these codes is compared to algebraic hard-in hard-out (HIHO) decoding and optimal trellis SISO decoding.  Binary antipodal signaling over AWGN channels is assumed throughout.

Initial parity-check matrices $H$ were formed by extending cyclic parity-check matrices for the respective $[31,21,5]$ and $[63,51,5]$ BCH codes \cite{MaSl78}.  These initial parity-check matrices were used as inputs to Algorithm \ref{tanner_graph_alg}, yielding  the parity-check matrices $H^\prime$, which in turn were used as inputs to Algorithm \ref{gtg_alg}, yielding $4$-cycle-free generalized Tanner graphs.  The suboptimal decoding algorithms implied by these graphical models are labeled $\TG(H)$, $\TG(H^\prime)$, and $\GTG(H^\prime)$, respectively.  The generalized Tanner graphs extracted for the $[32,21,6]$ and $[64,51,6]$ codes correspond to degree-$17$ and degree-$40$ generalized extensions, respectively.  Finally, the parity-check matrices $H^\prime$ were used as inputs to Algorithm \ref{twom_alg} with various values of $m^{\star}$.  The number $4$-, $6$-, and $8$-cycles ($N_4$, $N_6$, $N_8$) contained in the extracted graphical models for the $[32,21,6]$ and $[64,51,6]$ codes are given in Tables \ref{bch_32_21_table} and \ref{bch_64_51_table}, respectively.

The utility of Algorithm \ref{tanner_graph_alg} is illustrated in both Figures \ref{bch_32_21_fig} and \ref{bch_64_51_fig}: the $\TG(H^\prime)$ algorithms outperform the $\TG(H)$ algorithms by approximately $0.1$ dB and $0.5$ dB at a bit error rate (BER) of $10^{-4}$ for the $[32,21,6]$ and $[64,51,6]$ codes, respectively.  For both codes, the $4$-cycle-free generalized Tanner graph decoding algorithms outperform Tanner graph decoding by approximately $0.2$ dB at a BER of $10^{-4}$.  Further performance improvements are achieved for both codes by going beyond binary models.  Specifically, at a BER of $10^{-5}$, the suboptimal SISO decoding algorithm implied by the extracted $16$-ary graphical model for the $[32,21,6]$ code outperforms algebraic HIHO decoding by approximately $1.5$ dB.  The minimal trellis for this code is known to contain state variables with alphabet size at least $1024$ \cite{LaVa95}, yet the $16$-ary suboptimal SISO decoder performs only $0.7$ dB worse at a BER of $10^{-5}$.  At a BER of $10^{-4}$, the suboptimal SISO decoding algorithm implied by the extracted $32$-ary graphical model for the $[64,51,6]$ code outperforms algebraic HIHO decoding by approximately $1.2$ dB.  The minimal trellis for this code is known to contain state variables with alphabet size at least $4096$ \cite{LaVa95}; that a $32$-ary suboptimal SISO decoder loses only $0.7$ dB with respect to the optimal SISO decoder at a BER of $10^{-4}$ is notable.

\begin{table}[t]
\begin{center}
\begin{tabular}{c||c|c|c}
&$N_4$&$N_6$&$N_8$\\
\hline\hline
$\TG(H)$&$1128$&$37404$&$1126372$\\
\hline
$\TG(H^\prime)$ &$453$&$11152$&$260170$\\
\hline
$\GTG(H^\prime)$ &$0$&$62$&$298$\\
\hline
$4$-ary GM&$244$&$3852$&$50207$\\
\hline
$16$-ary GM&$70$&$340$&$724$
\end{tabular}
\caption{Short cycle structure of the initial and extracted graphical models for the $[32,21,6]$ extended BCH code.}\label{bch_32_21_table}
\end{center}
\end{table}

\begin{table}[t]
\begin{center}
\begin{tabular}{c||c|c|c}
&$N_4$&$N_6$&$N_8$\\
\hline\hline
$\TG(H)$&$9827$&$1057248$&$111375740$\\
\hline
$\TG(H^\prime)$&$3797$&$270554$&$19374579$\\
\hline
$\GTG(H^\prime)$&$0$&$163$&\small$1229$\\
\hline
$8$-ary GM&$847$&$19590$&$304416$\\
\hline
$32$-ary GM&$201$&$1384$&$0$
\end{tabular}
\caption{Short cycle structure of the initial and extracted graphical models for the $[64,51,6]$ extended BCH code.}\label{bch_64_51_table}
\end{center}
\end{table}
\section{Conclusion and Future Work}\label{conc_sec}
This work studied the space of graphical models for a given code in order to lay out some of the foundations of the theory of extractive graphical modeling problems.  The primary contributions of this work were the introduction of a new bound characterizing the tradeoff between cyclic topology and complexity in graphical models for linear codes and the introduction of a set of basic graphical model transformation operations which were shown to span the space of all graphical models for a given code.  It was demonstrated that these operations can be used to extract novel cyclic graphical models - and thus novel suboptimal  iterative soft-in soft-out (SISO) decoding algorithms - for linear block codes.

There are a number of interesting directions for future work motivated by the statement of the tree-inducing cut-set bound (TI-CSB).  While the minimal trellis complexity $s(\mathcal{C})$ of linear codes is well-understood, less is known about minimal tree complexity $t(\mathcal{C})$ and characterizing those codes for which $t(\mathcal{C})<s(\mathcal{C})$ is an open problem.  A particularly interesting open problem is the use of the Cut-Set Bound to establish an upper bound on the difference between $s(\mathcal{C})$ and $t(\mathcal{C})$; such a bound would allow for a re-expression of the TI-CSB in terms of the more familiar minimal trellis complexity.  A study of those codes which meet or approach the TI-CSB is also an interesting direction for future work which may provide insight into construction techniques for good codes with short block lengths (e.g. $10$s to $100$s of bits) defined on graphs with a few cycles (e.g. $3$, $6$ or $10$).  The development of statements similar to the TI-CSB for alternative measures of graphical model complexity and for graphical models of more general systems (e.g. group codes, nonlinear codes) is also interesting.

There are also a number of interesting directions for future work motivated by the study of graphical model transformation.  While the extracted graphical models presented in Section \ref{sim_results_subsec} are notable, ad-hoc techniques utilizing massively redundant models and judicious message filtering outperform the models presented in this work \cite{JiNa04,HaCh06c}.  Such massively redundant models contain many more short cycles than the models presented in Section \ref{sim_results_subsec} indicating that short cycle structure alone is not a sufficiently meaningful cost measure for graphical model extraction.  It is known that redundancy can be used to remove pseudocodewords (cf. \cite{VoKo05}) thus motivating the study of cost measures which consider both short cycle structure and pseudocodeword spectrum.  Finally, it would be interesting to study extraction heuristics beyond simple greedy searches, as well as those which use all of the basic graphical model operations (rather than just constraint merging).
\appendix\label{basic_move_appendix}
This appendix provides detailed definitions of both the $q^m$-ary graphical model properties described in Section \ref{model_prop_subsec} and the basic graphical model operations introduced in Section \ref{model_tx_sec}.  The proof of Lemma \ref{tree_comp_growth_lemma} is also further illustrated by example.  In order to elucidate these properties and definitions, a single-cycle graphical model for the extended Hamming code is studied throughout.  
\subsection{Single-Cycle Model for the Extended Hamming Code}
\begin{figure}[b]
\begin{center}
\includegraphics[width=2.25in]{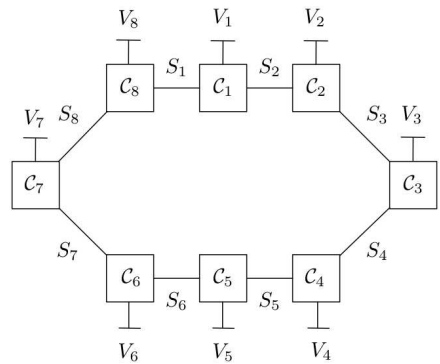}
\caption{Tail-biting trellis graphical model for the length $8$ extended Hamming code $\mathcal{C}_H$.}
\label{rm13_tbt_fig}
\end{center}
\end{figure}

Figure \ref{rm13_tbt_fig} illustrates a single-cycle graphical model (i.e. a tail-biting trellis) for the length $8$ extended Hamming code $\mathcal{C}_H$.
The hidden variables $S_1$ and $S_5$ are binary while $S_2$, $S_3$, $S_4$, $S_6$, $S_7$, and $S_8$ are $4$-ary.  All of the local constraint codes in this model are interface constraints.  Equations (\ref{tbt_gen_start})-(\ref{tbt_gen_end}) define the local constraint codes via generator matrices (where $G_i$ generates $\mathcal{C}_i$):
\setlength{\arraycolsep}{2.5pt}
\begin{align}\label{tbt_gen_start}
&\;\;\;\scriptstyle S_1\;V_1\;\:S_2&\scriptstyle S_2\;\:V_2\;\;S_3\;\;\;\:\nonumber\\
G_1=&
\left[\begin{array}{ccc}
1&0&10\\
0&1&01
\end{array}\right],
&G_2=
\left[\begin{array}{ccc}
10&1&10\\
01&1&01
\end{array}\right]\:
\end{align}
\begin{align}
&\;\;\;\;\scriptstyle S_3\;\;V_3\;\;S_4&\scriptstyle S_4\;\:V_4\;S_5\;\;\;\;\;\:\nonumber\\
G_3=&
\left[\begin{array}{ccc}
10&0&01\\
01&0&11\\
00&1&01
\end{array}\right],
&G_4=
\left[\begin{array}{ccc}
10&1&0\\
01&1&1
\end{array}\right]\;\;\;
\end{align}
\begin{align}
&\;\;\;\scriptstyle S_5\;V_5\;\:S_6&\scriptstyle S_6\;\:V_6\;\;S_7\;\;\;\:\nonumber\\
G_5=&
\left[\begin{array}{ccc}
1&0&10\\
0&1&01
\end{array}\right],
&G_6=
\left[\begin{array}{ccc}
10&1&10\\
01&1&01
\end{array}\right]\:
\end{align}
\begin{align}\label{tbt_gen_end}
&\;\;\;\;\scriptstyle S_7\;\;V_7\;\;S_8&\scriptstyle S_8\;\:V_8\;S_1\;\;\;\;\;\;\nonumber\\
G_7=&
\left[\begin{array}{ccc}
10&0&01\\
01&0&11\\
00&1&01
\end{array}\right],
&G_8=
\left[\begin{array}{ccc}
10&1&0\\
01&1&1
\end{array}\right].\;\;
\end{align}
The graphical model for $\mathcal{C}_H$ illustrated in Figure \ref{rm13_tbt_fig} is $4$-ary (i.e. $q=2$, $m=2$): the maximum hidden variable alphabet index set size is $2$ and all local constraints satisfy $\min\left(k(\mathcal{C}_i),n(\mathcal{C}_i)-k(\mathcal{C}_i)\right)\leq 2$.  The behavior, $\mathfrak{B}_H$, of this graphical model is generated by
\begin{align}
&\;\;\;\:\scriptstyle S_1\:V_1\;S_2\;\:V_2\;S_3\;\;V_3\;\;S_4\;V_4\;S_5\;V_5\;S_6\;\;V_6\;\:S_7\;V_7\;\;S_8\;\:V_8\nonumber\\
G_{\mathfrak{B}_H}=&\left[\begin{array}{cccccccccccccccc}
0&1&01&1&01&1&10&1&0&0&00&0&00&0&00&0\\
0&0&00&0&00&1&01&1&1&0&10&1&10&1&00&0\\
0&0&00&0&00&0&00&0&0&1&01&1&01&1&10&1\\
1&0&10&1&10&1&00&0&0&0&00&0&00&1&01&1
\end{array}\right].
\end{align}
The projection of $\mathfrak{B}_H$ onto the visible variable index set $I$, $\mathfrak{B}_{H|I}$, is thus generated by
\setlength{\arraycolsep}{4pt}
\begin{align}\label{cH_gen_mx}
&\scriptstyle \;\;\;V_1\;\;V_2\;\;V_3\;\;V_4\;\;V_5\;\;V_6\;\;V_7\;\;V_8\nonumber\\
G_{\mathfrak{B}_{H|I}}=&\begin{bmatrix}
1&1&1&1&0&0&0&0\\
0&0&1&1&0&1&1&0\\
0&0&0&0&1&1&1&1\\
0&1&1&0&0&0&1&1
\end{bmatrix}
\end{align}
which coincides precisely with a generator matrix for $\mathcal{C}_H$.\setlength{\arraycolsep}{5pt}
\subsection{$q^m$-ary Graphical Model Properties}
The three properties of $q^m$-ary graphical models introduced in Section \ref{model_prop_subsec} are discussed in detail in the following where it is assumed that a $q^m$-ary graphical model $\mathcal{G_C}$ with behavior $\mathfrak{B}$ for a linear code $\mathcal{C}$ over $\mathbb{F}_q$ defined on an index set $I$ is given.
\subsubsection{Internal Local Constraint Involvement Property}
Suppose there exists some hidden variable $S_j$ (involved in the local constraints $\mathcal{C}_{j_1}$ and $\mathcal{C}_{j_2}$) that does not satisfy the local constraint involvement property.  A new hidden variable $S_i$ that is a copy of $S_j$ is introduced to $\mathcal{G_C}$ by first redefining $\mathcal{C}_{j_2}$ over $S_i$ and then inserting a local repetition constraint $\mathcal{C}_i$ that enforces $S_j=S_i$.  The insertion of $S_i$ and $\mathcal{C}_i$ does not fundamentally alter the complexity of $\mathcal{G_C}$ since $n(\mathcal{C}_i)-k(\mathcal{C}_i)=|T_j|\leq m$ and since degree-$2$ repetition constraints are trivial from a decoding complexity viewpoint.  Furthermore, the insertion of $S_i$ and $\mathcal{C}_i$ does not fundamentally alter the cyclic topology of $\mathcal{G_C}$ since no new cycles can be introduced by this procedure.

\begin{figure}[htbp]
\begin{center}
\includegraphics[width=3.0in]{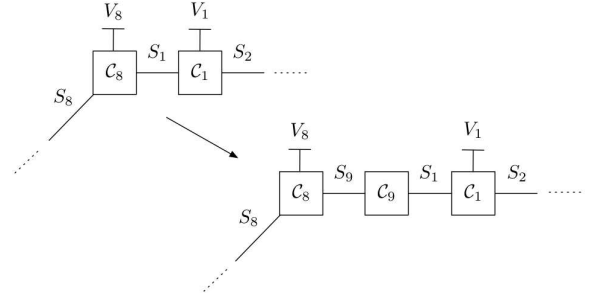}
\caption{Insertion of hidden variable $S_9$ and internal local constraint $\mathcal{C}_9$ into the tail-biting trellis for $\mathcal{C}_H$.}
\label{rm13_internal_fig}
\end{center}
\end{figure}

As an example, consider the binary hidden variable $S_1$ in Figure \ref{rm13_tbt_fig} which is incident on the interface constraints $\mathcal{C}_1$ and $\mathcal{C}_8$.  By introducing the new binary hidden variable $S_9$ and binary repetition constraint $\mathcal{C}_9$, as illustrated in Figure \ref{rm13_internal_fig}, $S_1$ can be made to be incident on the internal constraint $\mathcal{C}_9$.
The insertion of $S_9$ and $\mathcal{C}_9$ redefines $\mathcal{C}_8$ over $S_9$ resulting in the generator matrices
\setlength{\arraycolsep}{2.5pt}
\begin{align}
&\;\;\;\;\scriptstyle S_8\;V_8\;S_9&\scriptstyle S_9\;S_1\;\;\;\nonumber\\
G_8=&
\left[\begin{array}{ccc}
10&1&0\\
01&1&1
\end{array}\right],
&G_9=
\left[\begin{array}{cc}
1&1
\end{array}\right].
\end{align}
Clearly, the modified local constraints $\mathcal{C}_8$ and $\mathcal{C}_9$ satisfy the condition for inclusion in a $4$-ary graphical model.
\subsubsection{Internal Local Constraint Removal Property}
The removal of the internal constraint $\mathcal{C}_r$ from $\mathcal{G_C}$ in order to define the new code $\mathcal{C}^{\setminus r}$ proceeds as follows.  Each hidden variable $S_i$, $i\in I_S(r)$, is first disconnected from $\mathcal{C}_r$ and connected to a new degree-$1$ internal constraint $\mathcal{C}_{i^\prime}$ which does not impose any constraint on the value of $S_i$ (since it is degree-$1$).  The local constraint $\mathcal{C}_r$ is then removed from the resulting graphical model yielding $\mathcal{G}_{\mathcal{C}^{\setminus r}}$ with behavior $\mathfrak{B}^{\setminus r}$.  The new code $\mathcal{C}^{\setminus r}$ is the projection of $\mathfrak{B}^{\setminus r}$ onto $I$.

\begin{figure}[htbp]
\begin{center}
\includegraphics[width=2.75in]{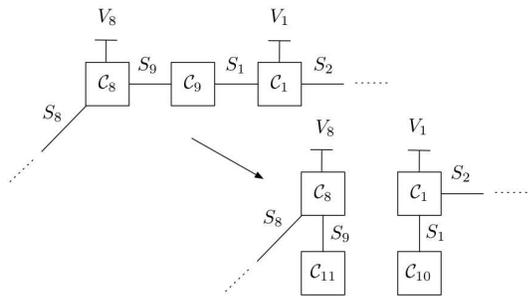}
\caption{Removal of internal local constraint $\mathcal{C}_9$ from the tail-biting trellis for $\mathcal{C}_H$.}
\label{rm13_removal_fig}
\end{center}
\end{figure}

As an example, consider the removal of the internal local constraint $\mathcal{C}_9$ from the graphical model for $\mathcal{C}_H$ described above; the resulting graphical model update is illustrated in Figure \ref{rm13_removal_fig}.  The new codes $\mathcal{C}_{10}$ and $\mathcal{C}_{11}$ are length 1, dimension 1 codes which thus impose no constraints on $S_{1}$ and $S_9$, respectively.
It is readily verified that the code $\mathcal{C}_{H}^{\setminus 9}$ which results from the removal of $\mathcal{C}_9$ from $\mathcal{C}_H$ has dimension $5$ and is generated by
\setlength{\arraycolsep}{4pt}
\begin{align}\label{G_Hminus9_def}
&\scriptstyle \;\;\;V_1\;\;V_2\;\;V_3\;\;V_4\;\;V_5\;\;V_6\;\;V_7\;\;V_8\nonumber\\
G_{H^{\setminus 9}}=&\begin{bmatrix}
1&1&1&1&0&0&0&0\\
0&0&1&1&0&1&1&0\\
0&0&0&0&1&1&1&1\\
0&1&1&0&0&0&1&1\\
1&0&1&0&0&1&1&0
\end{bmatrix}.
\end{align}
Note that $\mathcal{C}_{H}^{\setminus 9}$ corresponds to all paths in the tail-biting trellis representation of $\mathcal{C}_H$, not just those paths which begin and end in the same state.

The removal of an internal local constraint $\mathcal{C}_r$ results in the introduction of $|I_S(r)|$ new degree-$1$ local constraints.  Forney described such constraints as ``useless'' in \cite{Fo03} and they can indeed be removed from $\mathcal{G}_{\mathcal{C}^{\setminus r}}$ since they impose no constraints on the variables they involve.  Specifically, for each hidden variable $S_i$, $i\in I_S(r)$, involved in the (removed) local constraint $\mathcal{C}_r$, denote by $\mathcal{C}_j$ the other constraint involving $S_i$ in $\mathcal{G}_{\mathcal{C}}$.  The constraint $\mathcal{C}_j$ can be redefined as its projection onto $I_V(j)\cup \left\{I_S(j)\setminus i\right\}$.  It is readily verified that the resulting constraint $\mathcal{C}_j^{\setminus r}$ satisfies the condition for inclusion in a $q^m$-ary graphical model.

Continuing with the above example, $\mathcal{C}_{10}$, $\mathcal{C}_{11}$, $S_1$, and $S_9$  can be removed from the graphical model illustrated in Figure \ref{rm13_removal_fig} by redefining $\mathcal{C}_1$ and $\mathcal{C}_8$ with generator matrices
\setlength{\arraycolsep}{2.5pt}
\begin{align}
&\;\;\;\scriptstyle V_1\;S_2&\scriptstyle S_8\;V_8\;\;\;\;\;\;\nonumber\\
G_1=&
\left[\begin{array}{cc}
1&01\\
0&10
\end{array}\right],
&G_8=
\left[\begin{array}{cc}
10&1\\
01&1
\end{array}\right].\:
\end{align}
\subsubsection{Internal Local Constraint Redefinition Property}
Let $\mathcal{C}_i$ satisfy $n(\mathcal{C}_i)-k(\mathcal{C}_i)=m^\prime\leq m$ and consider a hidden variable $S_j$ involved in $\mathcal{C}_i$ (i.e. $j\in I_S(i)$) with alphabet index set $T_j$.  Each of the $\left|T_j\right|$ coordinates of $S_j$ can be redefined as a $q$-ary sum of some subset of the visible variable set as follows.  Consider the behavior $\mathfrak{B}^{\setminus i}$ and corresponding code $\mathcal{C}^{\setminus i}$ which result when $\mathcal{C}_i$ is removed from $\mathcal{G_C}$ (before $S_j$ is discarded).  The projection of $\mathfrak{B}^{\setminus i}$ onto $T_j\cup I$, $\mathfrak{B}^{\setminus i}_{|T_j\cup I}$, has length
\begin{equation}
n(\mathcal{C})+\left|T_j\right|
\end{equation}
and dimension
\begin{equation}
k(\mathcal{C}^{\setminus i})\geq k(\mathcal{C})
\end{equation}
over $\mathbb{F}_q$.  There exists a generator matrix for $\mathfrak{B}^{\setminus i}_{|T_j\cup I}$ that is systematic in some size $k(\mathcal{C}^{\setminus i})$ subset of the index set $I$ \cite{MaSl78}.  A parity-check matrix $H_j$ that is systematic in the $\left|T_j\right|$ positions corresponding to the coordinates of $S_j$ can thus be found for this projection; each coordinate of $S_j$ is defined as a $q$-ary sum of some subset of the visible variables by $H_j$.  Following this procedure, the internal local constraint $\mathcal{C}_i$ is redefined over $I$ by substituting the definitions of $S_j$ implied by $H_j$ for each $j\in I_S(i)$ into each of the $m^\prime$ $q$-ary single parity-check equations which determine $\mathcal{C}_i$.

Returning to the example of the tail-biting trellis for $\mathcal{C}_H$, the internal local constraint $\mathcal{C}_9$ is redefined over the visible variable set as follows.  The projection of $\mathfrak{B}_H^{\setminus 9}$ onto $T_1\cup I$ is generated by
\setlength{\arraycolsep}{4pt}
\begin{align}
&\scriptstyle \;\;\;S_1\;\;V_1\;\;V_2\;\;V_3\;\;V_4\;\;V_5\;\;V_6\;\;V_7\;\;V_8\nonumber\\
G_{\mathfrak{B}^{\setminus 9}_{H|T_1\cup I}}=&\begin{bmatrix}
0&1&1&1&1&0&0&0&0\\
0&0&0&1&1&0&1&1&0\\
0&0&0&0&0&1&1&1&1\\
1&0&1&1&0&0&0&1&1\\
1&1&0&1&0&0&1&1&0
\end{bmatrix}.
\end{align}
A valid parity-check matrix for this projection which is systematic in the position corresponding to $S_1$ is
\setlength{\arraycolsep}{4pt}
\begin{align}
&\scriptstyle \;\;\;S_1\;\;V_1\;\;V_2\;\;V_3\;\;V_4\;\;V_5\;\;V_6\;\;V_7\;\;V_8\nonumber\\
H_1=&\begin{bmatrix}
1&1&1&0&0&0&0&0&0\\
0&1&1&1&1&0&0&0&0\\
0&0&1&1&0&1&1&0&0\\
0&0&1&1&0&0&0&1&1
\end{bmatrix},
\end{align}
which defines the binary hidden variable $S_1$ as
\begin{equation}
S_1=V_1+V_2
\end{equation}
where addition is over $\mathbb{F}_2$.  A similar development defines the binary hidden variable $S_9$ as
\begin{equation}
S_9=V_5+V_8.
\end{equation}
The local constraint $\mathcal{C}_9$ thus can be redefined to enforce the single parity-check equation
\begin{equation}\label{C9_visible_def}
V_1+V_2+V_5+V_8=0.
\end{equation}

Finally, in order to illustrate the use of the $q^m$-ary graphical model properties in concert, denote by $\mathcal{C}_9^{(1)}$ the single parity-check constraint enforcing (\ref{C9_visible_def}).  It is readily verified that only the first four rows of $G_{H^{\setminus 9}}$ (as defined in (\ref{G_Hminus9_def})) satisfy $\mathcal{C}_9^{(1)}$.  It is precisely these four rows which generate $\mathcal{C}_H$ proving that
\begin{equation}\label{ch_ch9_eqn}
\mathcal{C}_H=\mathcal{C}_{H}^{\setminus 9}\cap\mathcal{C}_9^{(1)}.
\end{equation}
\subsection{Illustration of Proof of Lemma \ref{tree_comp_growth_lemma}}
\begin{figure}[htbp]
\begin{center}
\includegraphics[width=2.75in]{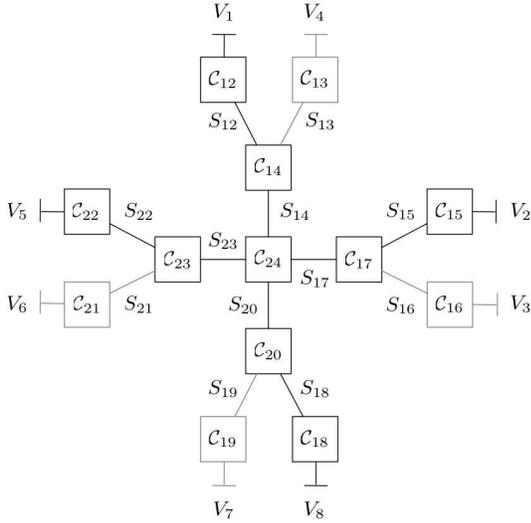}
\caption{Cycle-free binary graphical model for $\mathcal{C}_H^{\setminus 9}$.  The minimal spanning tree containing the interface constraints which involve $V_1$, $V_2$, $V_5$, and $V_8$, respectively, is highlighted.}
\label{rm13_binary_fig}
\end{center}
\end{figure}

In the following, the proof of Lemma \ref{tree_comp_growth_lemma} is illustrated by updating a cycle-free model for $\mathcal{C}_H^{\setminus 9}$ (as generated by (\ref{G_Hminus9_def})) with the single parity-check constraint defined by (\ref{C9_visible_def}) in order to obtain a cycle-free graphical model for $\mathcal{C}_H$.  A cycle-free binary graphical model for $\mathcal{C}_{H}^{\setminus 9}$ is illustrated in Figure \ref{rm13_binary_fig}\footnote{In order to emphasize that the code and hidden variable labels in Figure \ref{rm13_binary_fig} are in no way related to those labels used previously, the labeling of hidden variables and local constraints begin at $S_{12}$ and $\mathcal{C}_{12}$, respectively.}.  
All hidden variables in Figure \ref{rm13_binary_fig} are binary and the local constraints labeled $\mathcal{C}_{14}$, $\mathcal{C}_{17}$, $\mathcal{C}_{20}$, and $\mathcal{C}_{23}$ are binary single parity-check constraints while the remaining local constraints are repetition codes.  By construction, it has thus been shown that
\begin{equation}
t(\mathcal{C}_{H}^{\setminus 9})=1.
\end{equation}

In light of (\ref{C9_visible_def}) and (\ref{ch_ch9_eqn}), a $4$-ary graphical model for $\mathcal{C}_H$ can be constructed by updating the graphical model illustrated in Figure \ref{rm13_binary_fig} to enforce a single parity-check constraint on $V_1$, $V_2$, $V_5$, and $V_8$.  A natural choice for the root of the minimal spanning tree containing the interface constraints incident on these variables is $\mathcal{C}_{24}$.  The updating of the local constraints and hidden variables contained in this spanning tree proceeds as follows.  First note that since $\mathcal{C}_{12}$, $\mathcal{C}_{15}$, $\mathcal{C}_{18}$, and $\mathcal{C}_{22}$ simply enforce equality, neither these constraints, nor the hidden variables incident on these constraints, need  updating.  The hidden variables $S_{14}$, $S_{17}$, $S_{20}$, and $S_{23}$ are updated to be $4$-ary so that they send downstream to $\mathcal{C}_{24}$ the values of $V_1$, $V_2$, $V_8$, and $V_5$, respectively.  These hidden variable updates are accomplished by redefining the local constraints $\mathcal{C}_{14}$, $\mathcal{C}_{17}$, $\mathcal{C}_{20}$, and $\mathcal{C}_{23}$; the respective generator matrices for the redefined codes are
\setlength{\arraycolsep}{4pt}
 \begin{align}\label{4ary_redef_1}
&\;\;\;\scriptstyle S_{12}\;S_{13}\;S_{14}&\scriptstyle S_{15}\;S_{16}\;S_{17}\;\;\;\:\nonumber\\
G_{14}=&
\left[\begin{array}{ccc}
1&0&11\\
0&1&10
\end{array}\right],
&G_{17}=
\left[\begin{array}{ccc}
1&0&11\\
0&1&10
\end{array}\right]\:
\end{align}
 \begin{align}
&\;\;\;\scriptstyle S_{18}\;S_{19}\;S_{20}&\scriptstyle S_{21}\;S_{22}\;S_{23}\;\;\;\:\nonumber\\
G_{20}=&
\left[\begin{array}{ccc}
1&0&11\\
0&1&10
\end{array}\right],
&G_{23}=
\left[\begin{array}{ccc}
1&0&10\\
0&1&11
\end{array}\right]\:
\end{align}
Finally, $\mathcal{C}_{24}$ is updated to enforce both the original repetition constraint on the respective first coordinates of $S_{14}$, $S_{17}$, $S_{20}$, and $S_{23}$ and the additional single parity-check constraint on $V_1$, $V_2$, $V_5$, and $V_8$ (which correspond to the respective second coordinates of $S_{14}$, $S_{17}$, $S_{20}$, and $S_{23}$).  The generator matrix for the redefined $\mathcal{C}_{24}$ is
\setlength{\arraycolsep}{4pt}
\begin{align}\label{4ary_redef_2}
&\;\;\;\;\scriptstyle S_{14}\;\;\:S_{17}\;\;\:S_{20}\;\;\:S_{23}\nonumber\\
G_{24}=&
\left[\begin{array}{cccc}
10&10&10&10\\
01&00&00&01\\
00&01&00&01\\
00&00&01&01
\end{array}\right].
\end{align}

The updated constraints all satisfy the condition for inclusion in a $4$-ary graphical model.  Specifically, $\mathcal{C}_{24}$ can be decomposed into the Cartesian product of a length $4$ binary repetition code and a length $4$ binary single parity-check code.  The updated graphical model is $4$-ary and it has thus been shown by construction that
\begin{equation}
t(\mathcal{C}_H)\leq t(\mathcal{C}_{H}^{\setminus 9})+1=2.
\end{equation}
\subsection{Graphical Model Transformations}
The eight basic graphical model operations introduced in Section \ref{model_tx_sec} are discussed in detail in the following where it is assumed that a $q^m$-ary graphical model $\mathcal{G_C}$ with behavior $\mathfrak{B}$ for a linear code $\mathcal{C}$ over $\mathbb{F}_q$ defined on an index set $I$ is given.
\subsubsection{Local Constraint Merging}\label{constraint_merge_sec}\label{lc_merge_subsec}
\begin{figure}[h] 
\begin{center}
\includegraphics[width=2.75in]{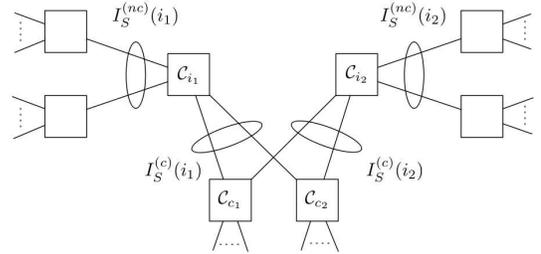}
\caption{Local constraint merging notation.  The local constraints $\mathcal{C}_{c_1}$ and $\mathcal{C}_{c_2}$ are common.}
\label{merging_variable_defs_fig}
\end{center}
\end{figure}

Suppose that two local constraints $\mathcal{C}_{i_1}$ and $\mathcal{C}_{i_2}$ are to be merged.  Without loss of generality, assume that there is no hidden variable incident on both $\mathcal{C}_{i_1}$ and $\mathcal{C}_{i_2}$ (since if there is, a degree-$2$ repetition constraint can be inserted).  The hidden variables incident on $\mathcal{C}_{i_1}$ may be partitioned into two sets
\begin{equation}
I_S(i_1)=I_S^{(c)}(i_1)\cup I_S^{(nc)}(i_1)
\end{equation}
where each $S_j$, $j\in I_S^{(c)}(i_1)$, is also incident on a constraint $\mathcal{C}_j$ which is adjacent to $\mathcal{C}_{i_2}$.  The hidden variables incident on $\mathcal{C}_{i_2}$ may be similarly partitioned.  The set of local constraints incident on hidden variables in both $I_S^{(c)}(i_1)$ and $I_S^{(c)}(i_2)$ are denoted \textit{common constraints} and indexed by $I_C^{(c)}(i_1,i_2)$.  Figure \ref{merging_variable_defs_fig} illustrates this notation. 

The merging of local constraints $\mathcal{C}_{i_1}$ and $\mathcal{C}_{i_2}$ proceeds as follows.  For each common local constraint $\mathcal{C}_j$, $j\in I_C^{(c)}(i_1,i_2)$, denote by $S_{j_1}$ ($S_{j_2}$) the hidden variable incident on $\mathcal{C}_j$ and $\mathcal{C}_{i_1}$ ($\mathcal{C}_{i_2}$).  Denote by $\mathcal{C}_{j|\{j_1,j_2\}}$ the projection of $\mathcal{C}_j$ onto the two variable index set $\{j_1,j_2\}$ and define a new $q^{k(\mathcal{C}_{j|\{j_1,j_2\}})}$-ary hidden variable $S_{j_1,j_2}$ which encapsulates the possible simultaneous values of $S_{j_1}$ and $S_{j_2}$ (as constrained by $\mathcal{C}_{j|\{j_1,j_2\}}$).  After defining such hidden variables for each $\mathcal{C}_j$, $j\in I_C^{(c)}(i_1,i_2)$, a set of new hidden variables results which is indexed by $I_S^{(c)}(i_1,i_2)$.  The local constraints $\mathcal{C}_{i_1}$ and $\mathcal{C}_{i_2}$ are then merged by replacing $\mathcal{C}_{i_1}$ and $\mathcal{C}_{i_2}$ by a code defined over
\begin{equation}
I_V(i_1)\cup I_V(i_2)\cup I_S^{(nc)}(i_1)\cup I_S^{(nc)}(i_2)\cup I_S^{(c)}(i_1,i_2)
\end{equation}
which is equivalent to $\mathcal{C}_{i_1}\cap\mathcal{C}_{i_2}$ and 
redefining each local constraint $\mathcal{C}_j$, $j\in I_C(i_1,i_2)$, over the appropriate hidden variables in $I_S^{(c)}(i_1,i_2)$.  

\begin{figure}[htbp]
\begin{center}
\includegraphics[width=2.25in]{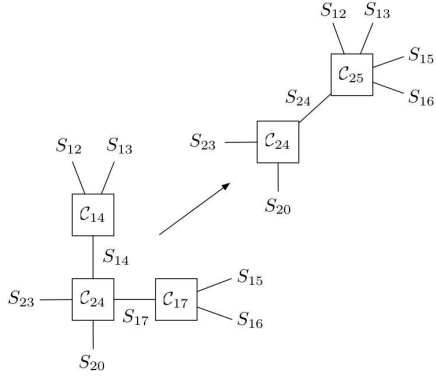}
\caption{The merging of constraints $\mathcal{C}_{14}$ and $\mathcal{C}_{17}$ in a $4$-ary graphical model for $\mathcal{C}_H$.  The resulting graphical model is $8$-ary.}
\label{constraint_merging_fig}
\end{center}
\end{figure}

As an example, consider again the $4$-ary cycle-free graphical model for $\mathcal{C}_H$ derived in the previous section, a portion of which is re-illustrated on the bottom left of Figure \ref{constraint_merging_fig}, and suppose that the local constraints $\mathcal{C}_{14}$ and $\mathcal{C}_{17}$ are to be merged.  The local constraints $\mathcal{C}_{14}$, $\mathcal{C}_{17}$, and $\mathcal{C}_{24}$ are defined by (\ref{4ary_redef_1}) and (\ref{4ary_redef_2}). 

The hidden variables incident on $\mathcal{C}_{14}$ are partitioned into the sets $I_S^{(c)}(14)=\{14\}$ and $I_S^{(nc)}(14)=\{12,13\}$.  Similarly, $I_S^{(c)}(17)=\{17\}$ and $I_S^{(nc)}(17)=\{15,16\}$.  The sole common constraint is thus $\mathcal{C}_{24}$.  The projection of $\mathcal{C}_{24}$ onto $S_{14}$ and $S_{17}$ has dimension $3$ and the new $8$-ary hidden variable $S_{25}$ is defined by the generator matrix
\setlength{\arraycolsep}{4pt}
 \begin{align}
&\;\;\;\;\;\scriptstyle S_{25}\;\;\;\:S_{14}\;\;\:S_{27}\nonumber\\
G_{14,17}=&
\left[\begin{array}{ccc}
100&10&10\\
010&01&00\\
001&00&01
\end{array}\right].
\end{align}
The local constraints $\mathcal{C}_{14}$ and $\mathcal{C}_{17}$ when defined over $S_{25}$ rather than $S_{14}$ and $S_{17}$, respectively, are generated by
\setlength{\arraycolsep}{4pt}
 \begin{align}
&\;\;\;\scriptstyle S_{12}\;S_{13}\;\;S_{25}&\scriptstyle S_{15}\;S_{16}\;\;S_{25}\;\;\;\;\;\;\;\;\nonumber\\
G_{14}^\prime=&
\left[\begin{array}{ccc}
1&0&110\\
1&0&111\\
0&1&100
\end{array}\right],
&G_{17}^\prime=
\left[\begin{array}{ccc}
1&0&101\\
1&0&111\\
0&1&100
\end{array}\right].\:
\end{align}
Finally, $\mathcal{C}_{24}$ is redefined over $S_{25}$ and generated by
\setlength{\arraycolsep}{4pt}
 \begin{align}
&\;\;\;\;\;\scriptstyle S_{25}\;\;\;\:S_{20}\;\;\:S_{23}\nonumber\\
G_{24}=&
\left[\begin{array}{ccc}
100&10&10\\
010&00&01\\
001&00&01\\
000&01&01
\end{array}\right]
\end{align}
while $\mathcal{C}_{14}$ and $\mathcal{C}_{17}$ are replaced by $\mathcal{C}_{25}$ which is equivalent to $\mathcal{C}_{14}\cap\mathcal{C}_{17}$ and is generated by
\setlength{\arraycolsep}{4pt}
 \begin{align}
&\;\;\;\;\;\scriptstyle S_{25}\;\:S_{12}\:S_{13}\:S_{15}\:S_{16}\;\nonumber\\
G_{25}=&
\left[\begin{array}{ccccc}
100&0&1&0&1\\
010&1&1&0&0\\
001&0&0&1&1
\end{array}\right].
\end{align}
Note that the graphical model which results from the merging of $\mathcal{C}_{14}$ and $\mathcal{C}_{17}$ is $8$-ary.  Specifically, $S_{24}$ is an $8$-ary hidden variable while $n(\mathcal{C}_{24})-k(\mathcal{C}_{24})=3$ and $k(\mathcal{C}_{25})=3$.
\subsubsection{Local Constraint Splitting}\label{lc_split_subsec}
Local constraint splitting is simply the inverse operation of local constraint merging.  Consider a local constraint $\mathcal{C}_j$ defined on the visible and hidden variables indexed by $I_V(j)$ and $I_S(j)$, respectively.  Suppose that $\mathcal{C}_j$ is to be split into two local constraints $\mathcal{C}_{j_1}$ and $\mathcal{C}_{j_2}$ defined on the index sets $I_V(j_1)\cup I_S(j_1)$ and $I_V(j_2)\cup I_S(j_2)$, respectively, such that $I_V(j_1)$ and $I_V(j_2)$ partition $I_V(j)$ while $I_S(j_1)\cup I_S(j_2)=I_S(j)$ but $I_S(j_1)$ and $I_S(j_2)$ need not be disjoint.  Denote by $I_S(j_1,j_2)$ the intersection of $I_S(j_1)$ and $I_S(j_2)$.  Local constraint splitting proceeds as follows.  For each $S_i$, $i\in I_S(j_1,j_2)$, make a copy $S_{i^\prime}$ of $S_i$ and redefine the local constraint incident on $S_i$ (which is not $\mathcal{C}_j$) over both $S_i$ and $S_{i^\prime}$.  Denote by $I_S^\prime(j_1,j_2)$ an index set for the copied hidden variables.  The local constraint $\mathcal{C}_j$ is then replaced by $\mathcal{C}_{j_1}$ and $\mathcal{C}_{j_2}$ such that $\mathcal{C}_{j_1}$ is defined over $I_V(j_1)\cup I_S(j_1)$ and $\mathcal{C}_{j_2}$  is defined over
\begin{equation}
I_V(j_2)\cup I_S(j_2)\setminus I_S(j_1,j_2)\cup I_S^\prime(j_1,j_2).
\end{equation}
Following this split procedure, some of the hidden variables in $I_S(j_1,j_2)$ and  $I_S^\prime(j_1,j_2)$ may have larger alphabets than necessary.  Specifically, if the dimension of the projection of $\mathcal{C}_{j_1}$ ($\mathcal{C}_{j_2}$) onto a variable $S_i$, $i\in I_S(j_1,j_2)$ ($i\in I_S^\prime(j_1,j_2)$), is smaller than the alphabet index set size of $S_i$, then $S_i$ can be redefined with an alphabet index set size equal to that dimension.

The merged code in the example of the previous section $\mathcal{C}_{25}$ can be split into two codes: $\mathcal{C}_{14}$ defined on $S_{12}$, $S_{13}$, and $S_{24}$, and $\mathcal{C}_{17}$ defined on $S_{15}$, $S_{16}$, and $S_{24^\prime}$.  The projection of $S_{24}$ onto $\mathcal{C}_{14}$ has dimension $2$ and $S_{24}$ can thus be replaced by the $4$-ary hidden variable $S_{14}$.  Similarly, the projection of $S_{24^\prime}$ onto $\mathcal{C}_{17}$ has dimension $2$ and $S_{24^\prime}$ can be replaced by the $4$-ary hidden variable $S_{17}$.
\subsubsection{Insertion/Removal of Degree-$2$ Repetition Constraints}\label{deg2_insert_subsec}
\begin{figure}[htbp]
\begin{center}
\subfigure[]{
\includegraphics[width=2.5in]{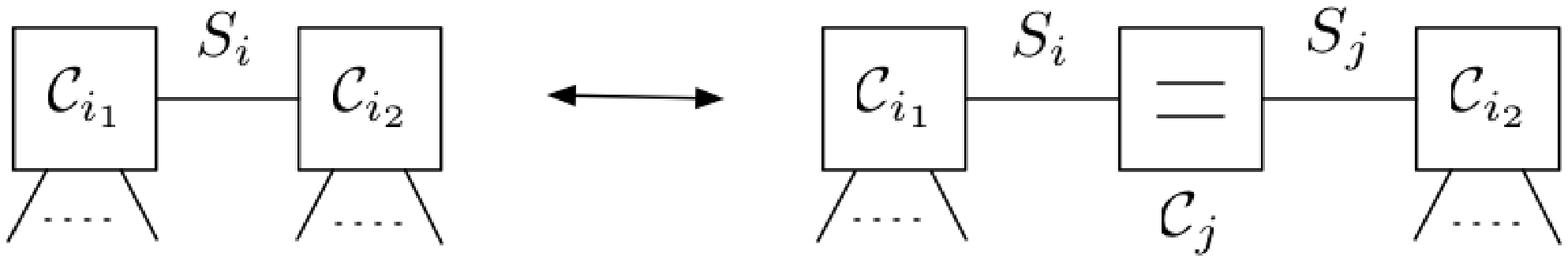}
\label{repetition_insertion_hidden_fig}}
\hspace{.4in}
\subfigure[]{\includegraphics[width=2.00in]{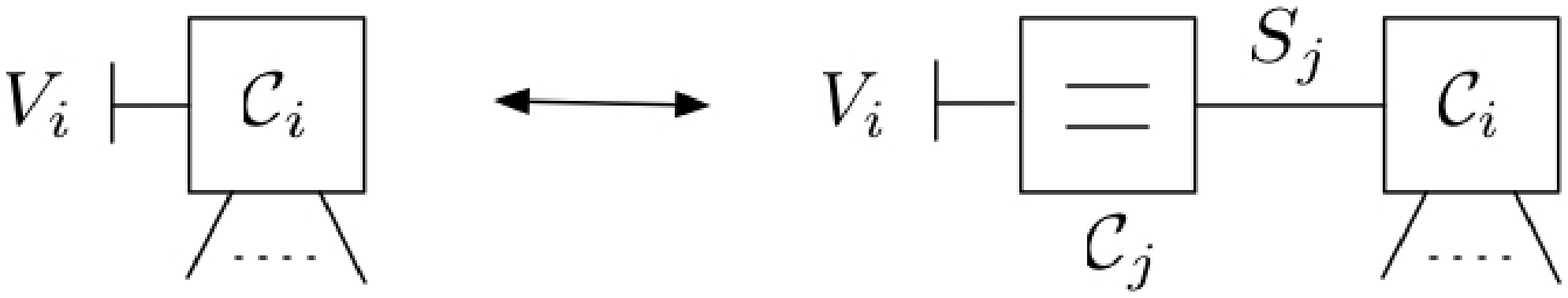}
\label{repetition_insertion_visible_fig}}
\caption{Insertion and removal of degree-$2$ repetition constraints.}\label{repetition_insertion_fig}
\end{center}
\end{figure}

Suppose that $S_i$ is a hidden variable involved in the local constraints $\mathcal{C}_{i_1}$ and $\mathcal{C}_{i_2}$.  A degree-$2$ repetition constraint is inserted by defining a new hidden variable $S_j$ as a copy of $S_i$, redefining $\mathcal{C}_{i_2}$ over $S_j$ and defining the repetition constraint $\mathcal{C}_j$ which enforces $S_i=S_j$.  Degree-$2$ repetition constraint insertion can be similarly defined for visible variables.  Conversely, suppose that $\mathcal{C}_j$ is a degree-$2$ repetition constraint incident on the hidden variables $S_i$ and $S_j$.  Since $\mathcal{C}_j$ simply enforces $S_i=S_j$, it can be removed and $S_j$ relabeled $S_i$.  Degree-$2$ repetition constraint removal can be similarly defined for visible variables.  The insertion and removal of degree-$2$ repetition constraints is illustrated in Figures \ref{repetition_insertion_hidden_fig} and \ref{repetition_insertion_visible_fig} for hidden and visible variables, respectively.
\subsubsection{Insertion/Removal of Trivial Constraints}\label{trivial_insert_subsec}
Trivial constraints are those incident on no hidden or visible variables so that their respective block lengths and dimensions are zero.  Trivial constraints can obviously be inserted or removed from graphical models.
\subsubsection{Insertion/Removal of Isolated Partial Parity-Check Constraints}\label{isolated_insert_subsec}
Suppose that $\mathcal{C}_{i_1},\ldots,\mathcal{C}_{i_j}$ are $j$ $q$-ary repetition constraints (that is each repetition constraint enforces equality on $q$-ary variables) and let $\beta_{i_1},\ldots,\beta_{i_j}\in\mathbb{F}_q$ be non-zero.  The insertion of an isolated partial parity-check constraint is defined as follows.  Define $j+1$ new $q$-ary hidden variables $S_{i_1},\ldots,S_{i_j}$ and $S_k$, and two new local constraints $\mathcal{C}_p$ and $\mathcal{C}_k$ such that $\mathcal{C}_p$ enforces the $q$-ary single parity-check equation
\begin{equation}
\sum_{l=1}^j\beta_{i_l}S_{i_l}=S_k
\end{equation}
and $\mathcal{C}_k$ is a degree-$1$ constraint incident only on $S_k$ with dimension $1$.  The new local constraint $\mathcal{C}_p$ defines the partial parity variable $S_k$ and is denoted \textit{isolated} since it is incident on a hidden variable which is involved in a degree-$1$, dimension $1$ local constraint (i.e. $\mathcal{C}_k$ does not constrain the value of $S_k$).  Since $\mathcal{C}_p$ is isolated, the graphical model that results from its insertion is indeed a valid model for $\mathcal{C}$.  Similarly, any such isolated partial parity-check constraint can be removed from a graphical model resulting in a valid model for $\mathcal{C}$.  

\begin{figure}[htbp]
\begin{center}
\includegraphics[width=2.75in]{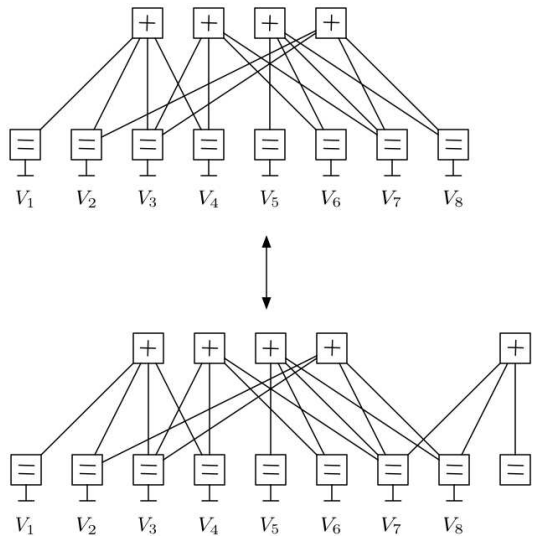}
\caption{The insertion/removal of an isolated partial parity-check constraint on $V_7$ and $V_8$ in a Tanner graph for $\mathcal{C}_H$.}
\label{isolated_insert_remove_fig}
\end{center}
\end{figure}

As an example, Figure \ref{isolated_insert_remove_fig} illustrates the insertion and removal of an isolated partial-parity check on the binary sum of $V_7$ and $V_8$ in a Tanner graph for $\mathcal{C}_H$ corresponding to (\ref{cH_gen_mx}) (note that $\mathcal{C}_H$ is self-dual so that the generator matrix defined in (\ref{cH_gen_mx}) is also a valid parity-check matrix for $\mathcal{C}_H$).


\end{document}